\documentclass[sigconf]{acmart}

\copyrightyear{2026}
\acmYear{2026}
\setcopyright{cc}
\setcctype{by}
\acmConference[CIKM '26]{Proceedings of the 35th ACM International Conference on Information and Knowledge Management}{November 07--11, 2026}{Rome, Italy}
\acmBooktitle{Proceedings of the 35th ACM International Conference on Information and Knowledge Management (CIKM '26), November 07--11, 2026, Rome, Italy}
\acmDOI{10.1145/3799682.3840904}
\acmISBN{979-8-4007-2539-5/2026/11}
 
\graphicspath{{./images/}} 

\usepackage{url}
\usepackage[show]{chato-notes}
\usepackage{tabularx}
\usepackage{multirow}
\usepackage{makecell}
\usepackage{calrsfs}
\usepackage{xcolor}
\usepackage{booktabs}
\usepackage{adjustbox}
\usepackage{caption}
\usepackage{subcaption}
\usepackage{tikz}
\usetikzlibrary{positioning}
\usepackage{mathtools}
\usepackage{amsmath}
\usepackage{soul}
\usepackage[most]{tcolorbox}
\usepackage[svgnames]{xcolor}

\usepackage{minted}

\usepackage{pifont}

\usepackage{enumitem} 

\usepackage{caption}
\newcommand{\uls}{\begin{itemize}[leftmargin=*]}
\newcommand{\ule}{\end{itemize}}
\newcommand{\ols}{\begin{enumerate}[leftmargin=*]}
\newcommand{\ole}{\end{enumerate}}
\newcommand{\li}{\item}

\newcommand{\para}[1]{\paragraph{\textnormal{\textbf{#1}}}} 

\usepackage{marginnote}

\usepackage{siunitx}
\DeclareMathAlphabet{\pazocal}{OMS}{zplm}{m}{n}
\DeclareMathAlphabet{\pazobfcal}{OMS}{cmsy}{b}{n}

\newcommand{\nb}[3]{
    {\colorbox{#2}{\bfseries\sffamily\scriptsize\textcolor{white}{#1}}}
    {\textcolor{#2}{$\blacktriangleright$\textsf\small{#3}$\blacktriangleleft$}}}

\newcommand{\ft}[1]{\nb{FT:}{blue}{#1}}

\newif\ifshowcomments
\showcommentsfalse 

\definecolor{darkgreen}{RGB}{0,100,0}

\newcommand{\structure}[1]{
    \ifshowcomments{\nb{Outline:}{red}{#1}}
    \fi
}
\newcommand{\commentCM}[1]{
    \ifshowcomments{\nb{Comment (Craig):}{purple}{#1}}
    \fi
}
\newcommand{\commentDG}[1]{
    \ifshowcomments{\nb{Comment (Debasis):}{darkgreen}{#1}}
    \fi
}

\usepackage{tikz}

\newcommand{\targetQ}{$q^\ast$}

\definecolor{pipelineblue}{RGB}{219,238,243}
\definecolor{predictiongreen}{RGB}{215,227,191}

\begin{document}


\title[Intermediate Answer State Prediction in Agentic RAG]{Predicting Partial Answer Quality and Utility in Agentic Retrieval-Augmented Generation}


\author{Fangzheng Tian}
\email{f.tian.1@research.gla.ac.uk}
\affiliation{%
  \institution{University of Glasgow}
  \city{Glasgow}
  \country{United Kingdom}
}

\author{Debasis Ganguly}
\email{Debasis.Ganguly@glasgow.ac.uk}
\affiliation{%
  \institution{University of Glasgow}
  \city{Glasgow}
  \country{United Kingdom}
}

\author{Craig Macdonald}
\email{Craig.Macdonald@glasgow.ac.uk}
\affiliation{%
  \institution{University of Glasgow}
  \city{Glasgow}
  \country{United Kingdom}
}


\renewcommand{\shortauthors}{Fangzheng Tian, Debasis Ganguly, and Craig Macdonald}


\begin{abstract}

Agentic Retrieval-Augmented Generation (RAG) has become a promising paradigm for multi-hop question answering, where a reasoning model iteratively issues queries to a retriever and incorporates newly retrieved context into subsequent reasoning steps.
While this iterative process can improve final answer quality, current evaluations of agentic RAG largely focus on end-to-end outcomes and provide limited visibility into how a model's answer state changes during generation. 
In this work, we introduce an in-trajectory probing framework to study intermediate answer states in agentic RAG. Specifically, after each retrieval-reasoning iteration, we force an agentic model to stop reasoning and generate an intermediate answer based on its current state.
This allows us to define two iteration-level measures: partial answer quality at each iteration, and partial utility as the change in partial answer quality across iterations.
Our analysis across multi-hop QA benchmarks reveals that partial answer quality often plateaus before natural termination, with many later iterations contributing only small measurable improvements.
Accordingly, we formulate two prediction tasks, partial answer quality prediction and partial utility prediction, and study trajectory-derived signals from intra-iteration, inter-iteration, and query-iteration perspectives.
Experiments show that partial answer quality is more predictable than partial utility, with supervised models achieving Pearson's $r$ above 0.43 for quality prediction.
Finally, using predicted answer quality and utility for early stopping reduces average iteration count by about 11\% while preserving about 98\% of the final answer quality achieved by natural stopping.

\end{abstract}



\maketitle

\section{Introduction}\label{s:introduction}


\begin{figure*}[tb]
\centering
\includegraphics[width=\textwidth]{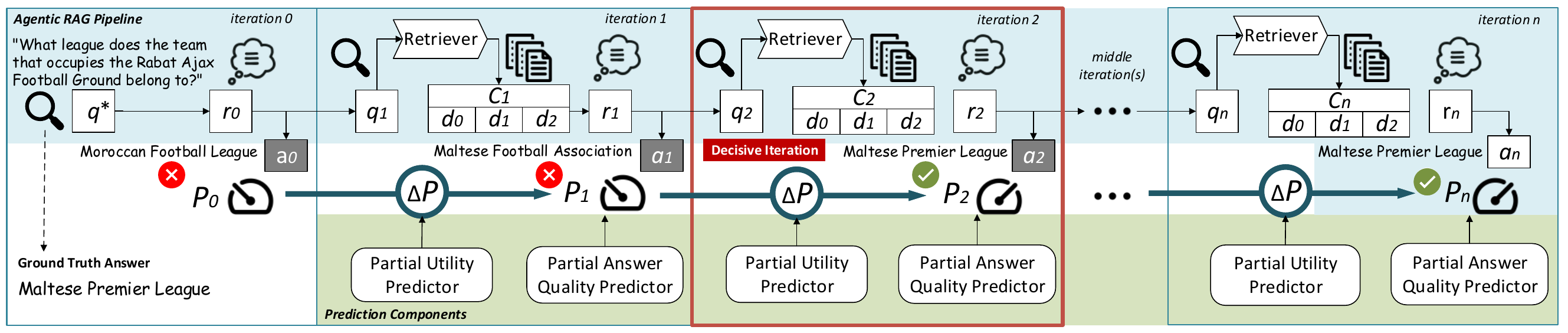}
\caption{Illustration of intermediate answer state analysis and prediction in agentic RAG. The \colorbox{pipelineblue}{blue} region shows the agentic RAG pipeline, where the model iteratively retrieves context $c_i$ with query $q_i$ and generates reasoning response $r_i$ before producing the final answer $a_n$.
After each iteration, in-trajectory probing obtains a probed answer $a_i$, whose partial answer quality is denoted as $P_i$.
The \colorbox{predictiongreen}{green} region shows two prediction targets: partial answer quality prediction, which estimates $P_i$, and partial utility prediction, which estimates the quality change $\Delta P$ between adjacent iterations.
In this example, iteration 2, highlighted in red, is a decisive iteration where the probed answer first becomes correct, after which the answer quality plateaus until the final answer.}
\label{fig:diagram}
\end{figure*}

Retrieval-Augmented Generation (RAG) has become a standard paradigm for question answering (QA) tasks~\cite{ragReview,ragInKnowledgeIntensiveNLP}. 
As user queries often involve complex information needs~\cite{surveyOnConversationalSearch}, such as multi-hop reasoning, agentic RAG has emerged as a rapidly developing direction~\cite{towardsAgenticRAGwithDeepReasoning}. 
Unlike conventional RAG, which typically performs a single retrieval step before answer generation, agentic RAG enables large language models (LLMs) to interact with retrievers iteratively throughout the reasoning process.
Figure~\ref{fig:diagram} illustrates this process in the blue region: starting from an input question, the model iteratively searches for additional evidence, reasons over retrieved information, generates follow-up queries, and eventually produces the final answer~\cite{searchR1,searchO1,coRAG}. 
Recent agentic RAG frameworks have shown improved effectiveness on benchmark QA tasks~\cite{2wiki,bamboogle,triviaQA,Musique,hotpotQA} compared with sequential RAG approaches~\cite{ragInKnowledgeIntensiveNLP,agenticRAGandQPP}.

\structure{Agentic is a black box, we don't know the contribution of individual iterations -> we conduct a quantitative study on it.} 

Despite their effectiveness, agentic RAG models remain largely opaque systems, as it is unclear how iterative retrieval and reasoning lead to answer-quality improvements~\cite{searchP1}.
Existing agentic RAG frameworks are commonly trained end-to-end for final answer correctness and format compliance, while the utility of each intermediate retrieval-reasoning iteration is rarely observed or explicitly supervised~\cite{searchR1,r1Searcher}.
Consequently, it remains unclear which iterations meaningfully improve the generation process and which merely incur additional computational cost.
Prior work has reported that agentic RAG can involve repeated iterations that prolong the process and degrade answer quality~\cite{agenticRAGandQPP}, suggesting that not all iterations are beneficial.

\commentDG{[first sentence] in terms of what? i don't follow why u say that this is a black box; i mean one can see the reasoning steps and the queries, right? \ft{I mean, the mechanism of how the reasoning and retrievals help improve RAG performance is a black box.} Is this work along similar lines to probing for explanations? does this contribute to increasing trustworthiness of RAG systems? can we also include some explanation literature in the related work section? \ft{Yes, they are reviewed in the second section of Related Work.}}

\commentCM{}

We argue that understanding which iterations are useful requires observing how the model's answer state changes throughout the agentic RAG process.
We therefore probe the model's answer after each iteration by prompting it to produce the most likely answer based on its current state, as illustrated by the grey answer boxes in Figure~\ref{fig:diagram}.
We refer to the quality of this probed answer as \emph{partial answer quality}, which measures how well the model would answer if the agentic process terminated at that iteration.
We further define \emph{partial utility} as the change in partial answer quality between adjacent iterations.
By tracking partial answer quality and partial utility across the trajectory, we can quantify the usefulness of each retrieval-reasoning iteration.

\structure{Define partial answer quality and partial utility. We find the decisive iterations are sparse and the answer quality often plateaus early.}

Through our study on two basic agentic RAG pipelines, we find that partial utility is sparsely distributed across iterations.
Most agentic RAG trajectories contain only one iteration that substantially changes partial answer quality, while many later iterations provide no measurable improvement.
For example, iteration 2 in Figure~\ref{fig:diagram} is a decisive iteration because it improves partial answer quality; afterwards, answer quality plateaus, and subsequent iterations no longer improve the answer.
This suggests substantial room for improving the efficiency of agentic RAG systems.

\commentDG{[about the sentence stating many iterations provide no improvements] little is informal and redundant; why don't u just say "small improvements"? \ft{I mean ``no'' improvements here}}

\commentCM{}

\structure{The room for improving the efficiency motivates us to predict the answer quality and utility.}

This opportunity to improve the efficiency of agentic RAG motivates us to predict intermediate answer states during the trajectory.
Such predictions can help identify when partial answer quality has plateaued and whether continuing the agentic process is likely to yield further improvement.
We formulate two prediction tasks: \emph{partial answer quality prediction} and \emph{partial utility prediction}.
The green region in Figure~\ref{fig:diagram} illustrates these two prediction modules.
Both predictions are made online during the agentic RAG process and use only the trajectory observed up to the current iteration.
We apply these predictions through an early-stopping strategy: the Agentic RAG trajectory can be terminated when the predicted answer quality remains consistently high, or when the predicted partial utility indicates that further iterations are unlikely to improve the answer.
In this way, the model can avoid redundant iterations and reduce computational cost while preserving final answer quality.

We argue that predicting partial answer quality and partial utility requires fine-grained analysis of the agentic RAG trajectory.
Unlike query performance prediction (QPP) for a retrieval step, prediction in RAG should account for how retrieved context affects the LLM's answer state for the target question~\cite{ragPredictions}.
This is more challenging in agentic RAG, where the utility of a later iteration depends on earlier retrieval and reasoning steps.
To address this, we organise predictive signals into three groups: (a) \textbf{intra-iteration signals}, which characterise retrieval and reasoning quality within a single iteration; (b) \textbf{inter-iteration signals}, which capture changes in information need, retrieved evidence, and reasoning across consecutive iterations; and (c) \textbf{query-iteration alignment signals}, which measure how each iteration relates to the original multi-hop question.
Based on this taxonomy, we examine unsupervised predictive signals and train supervised models using text inputs corresponding to the three signal groups.
We further study whether combining signals over a sliding window of recent iterations improves prediction effectiveness.
Finally, we apply the resulting predictions to simple heuristic early-stopping strategies to evaluate whether they can improve the efficiency of agentic RAG.

In summary, this paper makes three main contributions:
\ols
\li We introduce an in-trajectory probing framework for analysing intermediate answer states in agentic RAG. By probing the model after each retrieval-reasoning iteration, we quantify \emph{partial answer quality} and \emph{partial utility} along the trajectory.
\li We conduct a fine-grained analysis of intermediate answer states in basic agentic RAG models. Our results show that partial answer quality often plateaus before natural termination, and that only a small number of decisive iterations substantially change answer quality.
\li We formalise two prediction tasks for agentic RAG: \emph{partial answer quality prediction} and \emph{partial utility prediction}. We show that trajectory-derived signals can predict intermediate answer states and support early stopping, reducing the number of iterations with limited answer-quality loss.
\ole

The rest of the paper is organised as follows.
Section~\ref{s:related_work} reviews prior work on agentic RAG and RAG-related prediction tasks.
Section~\ref{s:methodology} defines partial answer quality and partial utility, formulates the prediction tasks, and introduces our prediction framework.
Section~\ref{s:early_stopping} introduces our prediction-guided early-stopping heuristics.
Section~\ref{s:experiments} presents the research questions and experimental setup.
Section~\ref{s:results} discusses the findings.
Finally, Section~\ref{s:conclusion} concludes the paper with limitations and future directions.

\section{Related Work}\label{s:related_work}

We review three lines of work related to our study: agentic RAG systems, probing methods for analysing intermediate model states, and prediction of retrieval utility in RAG.

\para{Agentic RAG}
Retrieval-Augmented Generation (RAG) augments language models with external evidence retrieved before generation~\cite{ragInKnowledgeIntensiveNLP}. 
Recent agentic RAG systems extend this retrieve-then-generate paradigm by allowing the model to interact with retrievers during reasoning, enabling adaptive, multi-step, or iterative retrieval~\cite{efficientRAG,iRCoT,kiRAG,RAT,FLARE,agenticRAGSurvey}. 
In these systems, the model can decide when additional evidence is needed, generate follow-up retrieval queries, and integrate newly retrieved information into its evolving reasoning state.

Existing agentic RAG methods differ in how retrieval is controlled. Early and prompting-based approaches, such as iRCoT~\cite{iRCoT} and FLARE~\cite{FLARE}, elicit retrieval actions during inference through interleaved reasoning and search. More recent systems, including Search-R1~\cite{searchR1} and R1-Searcher~\cite{r1Searcher}, train reasoning models with reinforcement learning to invoke retrieval during the reasoning process and improve final answer accuracy. 
Other work further augments the basic agentic RAG loop with additional control or refinement mechanisms. For example, Self-RAG~\cite{selfRAG} uses reflection tokens for adaptive retrieval, while TRACE~\cite{trace} builds knowledge-grounded reasoning chains for multi-hop QA.
Search-O1~\cite{searchO1} refines retrieved documents before injecting them into the reasoning chain, and CoRAG~\cite{coRAG} improves multi-step retrieval through dynamic query reformulation and path selection over candidate retrieval chains.

In this work, we study partial utility in the relatively simple agentic RAG setting, where the model follows a single retrieval-reasoning trajectory until producing the final answer. We instantiate this setting with two representative pipelines, Search-R1~\cite{searchR1} and R1-Searcher~\cite{r1Searcher}. This setting allows us to analyse iteration-level utility without the additional complexity introduced by modules such as self-reflection, document refinement or path selection.

\para{Probing intermediate answer states in RAG}
Our in-trajectory answer probing aims to elicit the model's intermediate answer state during the agentic RAG generation process. Existing work has analysed intermediate answer states in RAG mainly through uncertainty, internal representations, and self-evaluation signals.

Uncertainty-based methods estimate confidence from generated content, such as token-level uncertainty~\cite{FLARE} or sequence-level entropy~\cite{surveyOfHallucinations,detectHallucinations}. However, confidence is only a proxy for correctness, since LLMs may confidently hallucinate and retrieved contexts may introduce distracting or conflicting evidence~\cite{understandingIrrelevantContextInRAG,ragChecker}.
Internal-state methods further show that hidden representations can capture signals about knowledge use~\cite{ReDeEp}, grounding~\cite{FACTUM}, hallucination risk~\cite{internalStateForHallucinationDetection,llmInternalStateRevealHallucinationRisk}, or retrieval necessity~\cite{MeCo}. However, these methods typically require additional mappings from internal states to downstream outcomes and are not designed to directly measure iteration-level utility in agentic RAG.
LLM-based self-evaluation provides useful signals for estimating answer quality or retrieval necessity~\cite{llmKnowsWhatItKnows,selfRAG}, but such signals rely on the model's own judgment and may inherit overconfidence or hallucination tendencies~\cite{llmAreNotHumanLevelEvaluator}.

The closest work to ours is the probing method used in \cite{IGPO}, which measures intermediate answer states in agentic RAG by computing the model's probability of generating the ground-truth answer after each iteration. In contrast, our method elicits the model's most likely probed answer along the original trajectory and evaluates the generated answer against the ground truth. This directly measures intermediate answer quality and partial utility, and is easier to adapt to QA tasks with long-form answers~\cite{TREC-RAG-2025-overview,evalForTRECRAG2024}, where exact ground-truth likelihood is difficult to define.

\para{Predictions in RAG}
In RAG, answer quality is jointly affected by the LLM's internal knowledge and the retrieved context~\cite{ragRelevanceLabel}. Conversely, the utility of retrieved context is not determined solely by its lexical or semantic relevance to the input query, but by its effect on downstream generation quality~\cite{RelevanceAndUtility}. Useful retrieval results may be overlooked due to positional biases~\cite{lostInTheMiddle}, and their utility depends on the LLM's parametric knowledge, which may be sufficient, incomplete, or in conflict with the retrieved context~\cite{conflictsInRAG}. These factors make predicting RAG output quality and context utility more complicated than predicting the effectiveness of the LLM or retriever alone~\cite{ragPredictions}. 

A related line of work adapts query performance prediction (QPP) to predict context utility and answer quality in RAG. In traditional IR systems, QPP estimates retrieval effectiveness without relevance labels~\cite{QPP-goal}, using signals such as retrieval-score distributions~\cite{NQC,RSD,WRIG}, dense embeddings~\cite{denseQPP,aPairRatio}, or query-document interactions~\cite{bertQPP,ContextRichQPP,deepQPP}. Recent RAG-oriented prediction methods extend this idea to estimate whether retrieved context will improve generation quality~\cite{ragPredictions,RAGPredictionForQA}. These studies provide foundations for retrieval utility prediction, but they mainly focus on single-step RAG, where the retrieved context is fixed before generation.

Prediction becomes more challenging in agentic RAG because the retrieved context may affect not only the current answer state, but also subsequent reasoning, query formulation, and future retrieval decisions~\cite{iRCoT}. Therefore, utility and answer quality prediction in agentic RAG is a trajectory-level problem rather than a static prediction problem over a single query-context pair. Prior work has explored adaptive controllers for retrieval and stopping in iterative RAG~\cite{tcRAG,simRAG,AMBER,stopRAG}. For example, some methods estimate whether the accumulated evidence is sufficient to answer the question~\cite{simRAG,AMBER}, while Stop-RAG formulates stopping as a finite-horizon decision process~\cite{stopRAG}.

\para{Gap} Existing work provides limited insight into how answer quality evolves across retrieval-reasoning iterations.
Most studies focus on final answer quality or evidence sufficiency, rather than predicting intermediate answer states.
Our work differs in both target and methodology: instead of directly learning a stopping policy, we formulate partial answer quality and partial utility prediction as standalone tasks.
The closest work to ours is SIM-RAG~\cite{simRAG}, which uses an additional critic LLM to judge whether the accumulated retrieved context is sufficient.
In contrast, we combine interpretable unsupervised signals with supervised predictors derived from fine-grained trajectory analysis.
This makes our framework useful not only for prediction-guided stopping, but also for analysing how individual retrieval--reasoning iterations affect answer quality.

\section{Predictors for Agentic RAG}\label{s:methodology}

To support prediction-guided early stopping, this section defines partial answer quality and partial utility as measures of iteration-level contribution (Section \ref{ss:definition}), formulates their prediction tasks (Section \ref{ss:task}), and introduces our prediction scheme (Section \ref{ss:scheme}).

\subsection{Trajectories and Intermediate Answer States}\label{ss:definition}

\para{Agentic RAG trajectories}
Given an input question \targetQ{}, an agentic RAG system performs a sequence of retrieval-reasoning iterations before producing the final answer.
At iteration $i$, the generator formulates a query $q_i$, which is submitted to a retriever, and context $c_i$ is formed from the top-$k$ retrieved documents.
Conditioned on the input question, the previous trajectory, and the retrieved context in the current iteration, the generator then produces a reasoning response $r_i$.
After $n$ iterations, the system terminates the retrieval-reasoning process and yields the final answer $a_n$. The main part of Figure~\ref{fig:diagram} illustrates such a completed agentic RAG trajectory.

We denote the agentic RAG trajectory up to iteration $i$ as
\begin{equation}
    \tau_i = \left(q^\ast, r_0, \{q_j, c_j, r_j\}_{j=1}^{i}\right),
    \label{eq:trajectory}
\end{equation}
where $\tau_i$ is the trajectory prefix available at iteration $i$, $q^\ast$ is the input question, and $r_0$ denotes the initial reasoning response before external retrieval.
For each iteration $j$, $q_j$, $c_j$, and $r_j$ denote the generated query, the retrieved context, and the reasoning response, respectively.

\para{Intermediate answer} To observe the model's intermediate answer state, we probe the agentic RAG trajectory after each iteration.
Specifically, given the trajectory $\tau_i$ as defined in Equation~\eqref{eq:trajectory}, we prompt the generator to terminate the iterative process and produce an answer based on its current state as $a_i$.
In practice, we insert an answer-trigger token after iteration $i$ (which can be model-specific, e.g., Search-R1 uses ``\texttt{<answer>}'') to stop further iterations and use greedy decoding to obtain the most likely answer at the probing point.
This probed answer $a_i$ reflects the model's intermediate answer state for the input question conditioned on all the retrieved context and reasoning up to iteration $i$.
When $i=0$, the probed answer $a_0$ is solely determined by the model's parametric knowledge and initial reasoning.

\para{Partial answer quality and partial utility}
Given the ground-truth answer $a^\ast_q$ for \targetQ{}, we define partial answer quality $\pazocal{P}_i$ as the quality of the probed answer $a_i$ measured against $a^\ast_q$.
In our experiments, this quality is instantiated using F1.
The sequence $\{\pazocal{P}_0,\ldots,\pazocal{P}_n\}$ indicates how the model's answer quality evolves along the agentic RAG trajectory.

We further define the partial utility of iteration $i$ as the change in partial answer quality relative to the preceding probing point: $\pazocal{U}_i = \pazocal{P}_i - \pazocal{P}_{i-1}$, for $i \geq 1$.
Thus, $\pazocal{U}_i$ measures the incremental contribution of the $i$-th retrieval-reasoning iteration to the model's answer quality, where positive, negative, and near-zero values indicate improvement, degradation, and negligible change, respectively.

Importantly, while partial answer quality and partial utility can be measured when reference answers are available, they are not directly observable during inference. To use them for early stopping, we therefore formulate prediction tasks that estimate these metrics.

\subsection{Prediction Tasks}\label{ss:task}

\para{Partial Answer Quality Prediction}
Our first proposed prediction task estimates the quality of the model's intermediate answer after each iteration.
Given the trajectory $\tau_i$ observed up to iteration $i$ as defined in Equation~\eqref{eq:trajectory}, the goal of Partial Answer Quality Prediction is to predict the partial answer quality $\pazocal{P}_i$, without observing the ground-truth evaluation of the probed answer $a_i$.

For each iteration $i \geq 1$, a partial answer quality predictor maps the trajectory to a real-valued estimate of partial answer quality:
\begin{equation}
    \phi_{\pazocal{P}}: \tau_i \mapsto \mathbb{R}, (i\geq 1),
    \label{eq:answer_quality_prediction}
\end{equation}
where the input $\tau_i$ is the trajectory prefix observed up to iteration $i$, and the output is an estimate of $\pazocal{P}_i$.

Intuitively, an accurate answer quality predictor should assign higher scores to the iterations that yield better partial answers, and lower scores to those that yield worse ones. Hence, the estimates can be used to detect whether answer quality has plateaued.

\para{Partial Utility Prediction}
The second prediction task estimates how much an iteration changes the model's answer quality, i.e., the change in answer quality $\pazocal{P}$ caused by iteration $i$ compared with the previous iteration $i-1$.
Formally, the target of this task is the partial utility $\pazocal{U}_i$.

For each iteration $i \geq 1$, we define a partial utility predictor as
\begin{equation}
    \phi_{\pazocal{U}}: \tau_i \mapsto \mathbb{R}, (i\geq 1),
    \label{eq:utility_prediction}
\end{equation}
where the input $\tau_i$ is the trajectory prefix observed up to iteration $i$, and the output is an estimate of $\pazocal{U}_i$.

Since both prediction tasks take the trajectory prefix $\tau_i$ as input, the predictors need to extract signals from it that are informative for their respective target metrics. We next discuss the types of trajectory-derived signals and introduce our prediction scheme.

\newcommand{\tok}[1]{\texttt{#1}}
\newcommand{\dash}{--}

\begin{table*}[t]
\centering
\caption{
Overview of predictive signals for intermediate answer-state prediction.
For each signal source, we instantiate unsupervised features (left) and supervised cross-encoder regressors (right).
Unsupervised predictive signals are used as indirect indicators, while supervised predictors are trained directly on the corresponding target metric.
All predictive signals are computed from the trajectory prefix available at iteration $i$ and are used to predict partial answer quality $\pazocal{P}_i$ or partial utility $\pazocal{U}_i$.
The trajectory component notations follow the definitions in Section~\ref{ss:definition}; $\sigma$ denotes a similarity measurement.
}
\begin{adjustbox}{max width=\textwidth}
\begin{tabular}{l l l l l}
\toprule
\multirow{2}{*}{\textbf{Signal Source}}
& \multicolumn{3}{c}{\textbf{Unsupervised Predictive Signals}}
& \multicolumn{1}{c}{\textbf{Supervised Predictive Signals}} \\
\cmidrule(lr){2-4}\cmidrule(lr){5-5}
& \makecell{\textbf{Quality Estimator}}
& \textbf{Similarity}
& \makecell{\textbf{Answer Confidence}}
& \textbf{Cross-encoder Regressor} \\
\midrule

\textbf{Intra-iteration}
&
QPP$(q_i, c_i)$
&
\dash
&
$\mathrm{prob}(a_i)$
&
\tok{[CLS]} \targetQ \tok{[CURR\_Q]} $q_i$ \tok{[CURR\_R]} $r_i$ \tok{[CURR\_C]} $c_i$ \\


\textbf{Inter-iteration}
&
\dash
&
$\sigma(r_{i-1}, r_i)$; $\sigma(c_{i-1}, c_i)$
&
$\Delta \mathrm{prob}(a_i)$
&
\tok{[CLS]} \targetQ \tok{[PREV\_Q]} $q_{i-1}$ \tok{[PREV\_R]} $r_{i-1}$ \tok{[CURR\_Q]} $q_i$ \tok{[CURR\_R]} $r_i$ \\


\textbf{Question-Iteration}
&
\dash
&
$\sigma(q^\ast, q_i)$; $\sigma(r_0, r_i)$
&
\dash
&
\tok{[CLS]} \targetQ \tok{[ORIG\_R]} $r_0$ \tok{[CURR\_Q]} $q_i$ \tok{[CURR\_R]} $r_i$ \\

\bottomrule
\end{tabular}
\end{adjustbox}

\label{table:taxonomy}

\end{table*}

\subsection{Prediction Scheme}\label{ss:scheme}

\para{Source of Predictive Signals}
Agentic RAG trajectories are sequential: each iteration is influenced by the retrieval and reasoning steps that precede it.
Thus, predicting the intermediate answer state at iteration $i$ requires analysing not only the current iteration, such as its retrieval quality, but also its relation to previous iterations and its alignment with the original input question.
We therefore consider three types of predictive signals: intra-iteration signals, inter-iteration signals, and query-iteration alignment signals.
These three signal types correspond to the three rows in Table~\ref{table:taxonomy}.
We introduce each type and its intuition below.

\ols

\li \textbf{Intra-iteration analysis.} This category characterises the current iteration $i$ by analysing its internal components: query $q_i$, retrieved context $c_i$, reasoning response $r_i$, and probed answer $a_i$.
Our intuition is that a high-quality intermediate answer is more likely when the retrieved context is relevant, the reasoning is coherent, and the probed answer is confident.

\li \textbf{Inter-iteration changes.} This category captures how the trajectory evolves across iterations, such as changes in query information needs and reasoning responses.
Large changes may indicate that the current iteration introduces new useful information, shifts to a new information need, or substantially changes the answer state.
In contrast, small changes may suggest repeated retrieval, unresolved information needs, or plateaued answer quality.
These signals are therefore especially relevant to partial utility prediction.

\li \textbf{Question-iteration alignment.}
This category measures how the current iteration relates to the original input question \targetQ{}.
Since each query may address only part of a multi-hop information need, later iterations can either remain focused on the target question or drift toward redundant or irrelevant exploration.
We analyse this alignment by comparing the current query and reasoning response with the target question and initial reasoning.
A high alignment suggests that the iteration remains focused on evidence needed to answer \targetQ{}, while low alignment may indicate topic drift~\cite{informationNeed&Query&QPP}.
Although topic drift can sometimes reflect the exploration of a new aspect of the information need, it still provides a useful dimension for analysing the agentic RAG trajectory.

\ole

\para{Predictor Instantiations}

For each signal source, we instantiate predictors for partial answer quality and partial utility, as summarised in Table~\ref{table:taxonomy}.

The left side of Table~\ref{table:taxonomy} shows \emph{unsupervised predictive signals}, which are computed directly from the agentic RAG trajectory without task-specific training.
They capture interpretable aspects of the trajectory from the three signal sources.
For intra-iteration analysis, retrieval-quality predictors estimate whether the retrieved context is likely to support the current query.
For inter-iteration analysis, similarity or overlap between adjacent retrieval results and reasoning responses indicates how much the model's information need and reasoning state change from one iteration to the next.
For query-iteration alignment, similarity between the current iteration and the original question or initial reasoning indicates whether the trajectory remains focused on the target information need.
These predictors can use retrieval scores, document embeddings, retrieved-list overlap, and generated text representations.
Although unsupervised predictors provide interpretable indicators of the trajectory state, they are not directly optimised for partial answer quality $\pazocal{P}_i$ or partial utility $\pazocal{U}_i$.

The second form is \emph{supervised predictive signals}, which directly estimate partial answer quality $\pazocal{P}_i$ or partial utility $\pazocal{U}_i$.
We instantiate them as lightweight cross-encoder regressors over the structured textual inputs shown on the right side of Table~\ref{table:taxonomy}.
All inputs include the original question as a task anchor, while separately modelling intra-iteration, inter-iteration, or query-iteration relations.

\para{Sequential Combination}
The predictors above estimate the targets defined in Equations~\eqref{eq:answer_quality_prediction} and~\eqref{eq:utility_prediction} from individual signal sources.
However, agentic RAG trajectories are sequential, and the answer state at iteration $i$ may depend on the prediction patterns observed in previous iterations.
For example, stable high-quality predictions may indicate that answer quality has plateaued, while repeated low-utility predictions may suggest that further retrieval is unlikely to help~\cite{agenticRAGandQPP}.
Following prior work on multi-turn performance prediction, where evidence from previous turns is used to predict later outcomes~\cite{PredRetrFailInConvRec}, we combine predictor outputs in a sequential prediction scheme.

For each iteration $i$, we collect the outputs of all estimators listed in Table~\ref{table:taxonomy} from the most recent-iteration window.
Let $\mathbf{o}_j$ denote the vector of estimator outputs available at iteration $j$.
We define the windowed feature vector as
\begin{equation}
    \mathbf{z}_i =
    \left[
    \mathbf{o}_{i-w+1}\oplus
    \mathbf{o}_{i-w+2}\oplus
    \ldots\oplus
    \mathbf{o}_{i}
    \right],
    \label{eq:sequential_features}
\end{equation}
where $w$ is the window size and $\oplus$ denotes vector concatenation.
When $i < w$, the context window extends beyond the beginning of the trajectory and causes unavailable positions, so we pad these unavailable positions with missing-value indicators.

We train a lightweight prediction head to approximate the target metric from the windowed feature vector $z_i$.
The target is either partial answer quality $\pazocal{P}_i$ or partial utility $\pazocal{U}_i$.
The prediction head can be instantiated by a regression model, such as a linear regressor or an MLP regressor.
By combining estimator outputs across a recent-iteration window, the prediction head exploits complementary signals and temporal patterns in the trajectory, with the goal of improving prediction effectiveness over individual estimators.

\section{Prediction-Guided Early Stopping}
\label{s:early_stopping}

Using the prediction scheme introduced in Section~\ref{ss:scheme}, we further use predictions of partial answer quality and partial utility to guide early stopping in agentic RAG.
Our motivation is that a probed answer at iteration $i$ provides a candidate output if the system stops at that point.
If the predicted answer quality is high and the current iteration has positive predicted utility, the system can stop early and return the candidate output, since further iterations are unlikely to improve answer quality substantially.
Importantly, this stopping decision does not require probing at every iteration during inference: predictions can also be made from signals extracted from the main agentic RAG trajectory alone.

\para{Prediction-defined answer states} To make the stopping decision robust, we use the two predictions as complementary evidence.
The predicted partial answer quality $\hat{\pazocal{P}}_i$ estimates whether the current answer is good enough, while the predicted partial utility $\hat{\pazocal{U}}_i$ estimates whether the current iteration improves answer quality relative to the previous one.
Using either prediction alone can be unreliable: after a high predicted-utility iteration, answer quality may still be low, while low predicted utility may indicate either a plateau after reaching a good answer or stagnation at a poor answer.
We therefore combine the two predictions and stop only when they jointly provide a clear and consistent signal.
Since partial answer quality is instantiated as F1 in our experiments, the partial answer quality and partial utility satisfy $\pazocal{P}_i \in [0,1]$, and $\pazocal{U}_i \in [-1,1]$.

\begin{figure}[t]
\centering
\begin{tikzpicture}[scale=0.9, every node/.style={font=\small}]
    \draw[thick] (0,0) rectangle (2.8,2.8);

    \draw[dashed] (1.4,0) -- (1.4,2.8);
    \draw[dashed] (0,1.4) -- (2.8,1.4);

    \draw[->, thick] (-0,0) -- (3,0) node[right] {$\hat{\pazocal{P}}_i$};
    \draw[->, thick] (0,-0) -- (0,3) node[above] {$\hat{\pazocal{U}}_i$};

    \node[below] at (1.4,0) {$\theta_P$};
    \node[left] at (0,1.4) {$\theta_U$};

    \node[align=center] at (2.1,2.1) {\textbf{State 0}\\high $P$,\\high $U$};
    \node[align=center] at (2.1,0.7) {\textbf{State 1}\\high $P$,\\low $U$};
    \node[align=center] at (0.7,2.1) {\textbf{State 2}\\low $P$,\\high $U$};
    \node[align=center] at (0.7,0.7) {\textbf{State 3}\\low $P$,\\low $U$};
\end{tikzpicture}
\caption{Four states in the prediction-guided early-stopping heuristic.
The thresholds $\theta_P$ and $\theta_U$ partition the space of predicted partial answer quality $\hat{\pazocal{P}}_i$ and predicted partial utility $\hat{\pazocal{U}}_i$ into states 0--3.}
\label{fig:answer_states}
\end{figure}
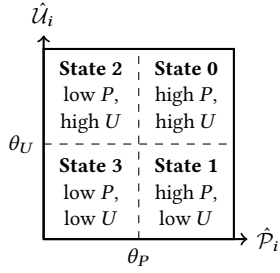

Figure~\ref{fig:answer_states} illustrates the four \emph{answer states} used by our prediction-guided early-stopping heuristic.
At each iteration $i$, the current answer state is assigned according to the predicted partial answer quality $\hat{\pazocal{P}}_i$ and predicted partial utility $\hat{\pazocal{U}}_i$.
The thresholds $\theta_P$ for predicted partial answer quality and $\theta_U$ for predicted partial utility are selected within the corresponding target ranges and together define four states.
State 0 corresponds to high answer quality and high utility; State 1 to high answer quality and low utility; State 2 to low answer quality and high utility; and State 3 to low answer quality and low utility.

\para{State-based stopping heuristics} Based on these states defined by predicted answer quality and utility, our heuristic follows two stopping rules.
First, if the current iteration reaches State 0, we stop immediately and return the current probed answer, since the system is predicted to have reached a high-quality and still-improving answer state.
Second, if the trajectory enters State 3 after previously reaching State 0 or State 1, we stop and return the earliest previous answer from State 0 or State 1. This \emph{rollback rule} is designed to maximise efficiency: among predicted high-quality states, we select the earliest one as the stopping output. For evaluation, however, the iteration cost is counted up to the point where the rollback decision is triggered, because the system must first observe the later State 3 before deciding to roll back.
For State 2 and all other cases, the agentic RAG process continues until natural termination, because the predictions do not yet provide a sufficiently clear signal to make early-stopping decisions.

\section{Experimental Setup}\label{s:experiments}

We introduce the research questions on analysing and predicting partial answer quality and utility (Section \ref{ss:rqs}), the implemented predictors (Section \ref{ss:implementations}), and the experimental setups (Section \ref{ss:setups}).

\subsection{Research Questions}\label{ss:rqs}

Our in-trajectory probing method introduced in Section~\ref{ss:definition} allows us to observe the model's intermediate answer state after each retrieval-reasoning iteration.
We use this observation to analyse how partial answer quality and partial utility change along agentic RAG trajectories.
Specifically, we examine whether answer quality improves steadily with additional iterations, or whether the agentic RAG process often continues after answer quality has already plateaued.
This leads to our first research question:

\uls
\li \textbf{RQ-1}: How do partial answer quality and partial utility evolve across agentic RAG trajectories?
\ule

After analysing the evolution of intermediate answer states, we examine whether partial answer quality and partial utility can be predicted from trajectory-derived signals, using the prediction methodology introduced in Section~\ref{ss:scheme}.
Therefore, we formulate the second research question as:

\uls
\li \textbf{RQ-2}: Can we predict partial answer quality and partial utility?
\ule

Finally, we evaluate whether predictions can support process control in agentic RAG.
Since redundant iterations increase computational cost, a useful predictor should identify when the system can stop with limited answer-quality loss.
We therefore conduct a simple early-stopping experiment guided by predicted intermediate answer states, leading to our final research question:

\uls
\li \textbf{RQ-3}: Can intermediate answer state prediction improve the efficiency of agentic RAG systems?
\ule

\subsection{Implemented Predictors}\label{ss:implementations}

In the experiments, we instantiate the predictor scheme introduced in Section~\ref{ss:scheme} and summarised in Table~\ref{table:taxonomy}.
The implemented predictors follow the three trajectory relations defined earlier: intra-iteration analysis, inter-iteration change, and question-iteration alignment.
For each relation, we consider unsupervised measurements and supervised cross-encoder regressors.
We additionally use answer-confidence signals as optional predictors; however, these signals require answer probing at inference time and therefore introduce extra computational overhead.
We describe the implemented predictors below.

\ols
\li \textbf{Unsupervised methods (Unsup.)}: For intra-iteration signals, we implement QPP methods as retrieval-quality predictors based on retrieval-score variation (NQC~\cite{NQC}) and document-embedding structure, including semantic concentration~\cite{denseQPP} and pairwise document coherence~\cite{aPairRatio}.
For inter-iteration signals, we compute rank-biased overlap (RBO)~\cite{RBO} between adjacent retrieval lists and SBERT cosine similarity between adjacent reasoning responses.
For question-iteration alignment, we compute RBO between the retrieval list of the original question and that of the current query, as well as SBERT cosine similarity between the initial reasoning response $r_0$ and the current reasoning response $r_i$.
\li \textbf{Supervised methods (Sup.)}: Following the input formats in Table~\ref{table:taxonomy}, we fine-tune an SBERT-based cross-encoder~\cite{SBERT}\footnote{\texttt{sentence-transformers/all-MiniLM-L6-v2}} as a regressor. Each regressor takes the trajectory components available at the current iteration as input and predicts either partial answer quality or partial utility. All regressors are trained with mean squared error (MSE) loss.
\li \textbf{Probing-based methods (Prob.)}: We use the length-normalised generation probability of the probed answer as a confidence signal. As an intra-iteration signal, this probability estimates how confidently the model generates the answer at the current probing point. As an inter-iteration signal, the difference in answer generation probability between adjacent probing points captures changes in the model's confidence along the trajectory.
\ole

For sequential prediction, we combine signals from recent iterations using a lightweight multilayer perceptron (MLP).
For each target iteration, we concatenate the predictive signals from a fixed-length history window and use the resulting vector as the MLP input.
The MLP has two hidden layers with 16 and 8 hidden units.
We vary the window size from 1 to 5; $w=1$ uses only signals from the current iteration, while larger windows include predictions made for previous iterations.

For the early-stopping experiment, we apply the heuristic introduced in Section~\ref{s:early_stopping} using the best-performing predictor combinations.
We sweep the quality and utility thresholds ($\theta_P$ and $\theta_U$) over their respective value ranges to obtain different operating points, and evaluate the trade-off between average iteration count and final answer quality.
As reference baselines, we use fixed-cap stopping, denoted as Cap@$k$, where the agentic RAG process is terminated after at most $k$ iterations.

Since partial answer quality prediction and partial utility prediction in agentic RAG have not been studied as stand-alone tasks, there are no established task-specific baselines.
We therefore evaluate a spectrum of predictors that differ in supervision, interpretability, and inference-time cost.
Training-free unsupervised predictors provide diagnostic baselines, supervised predictors test direct optimisation for the proposed targets, and probing-based predictors assess the predictive value of confidence signals from probed answers.
This setup allows us to compare different predictor families under the same evaluation protocol.

We further evaluate different combinations of predictor families and context-window sizes as ablations.
These ablations allow us to assess the contribution of each predictor family and whether incorporating previous iterations improves prediction effectiveness.

\subsection{Agentic RAG Pipelines and Datasets}\label{ss:setups}

\para{Agentic RAG Configurations.}
We study two representative RL-trained agentic RAG pipelines built on Qwen2.5-7B~\cite{qwen25}: Search-R1~\cite{searchR1} and R1-Searcher~\cite{r1Searcher}.
Since our goal is to analyse how intermediate answer quality evolves across retrieval-reasoning iterations, relying on a single pipeline may conflate our findings with the behaviour of one specific training setup.
Using two systems allows us to examine whether the observed patterns are shared across RL-trained agentic RAG models, while still keeping the setting controlled.
We focus on the basic iterative paradigm rather than variants with additional optimisation modules, so that we can analyse the underlying dynamics of agentic RAG itself, especially how partial answer quality changes along the trajectory and how the model's self-decided stopping behaviour emerges.

\ols
\li \textbf{Search-R1}~\cite{searchR1} trains the reasoning model with single-stage outcome-based RL on HotpotQA~\cite{hotpotQA} and NQ~\cite{NQ}.
It enables the model to invoke retrieval during reasoning and incorporate retrieved evidence into the reasoning trace.
\li \textbf{R1-Searcher}~\cite{r1Searcher} trains the reasoning model with a two-stage outcome-based RL procedure on HotpotQA and 2WikiMultiHopQA.
Compared with Search-R1, it is trained with two explicit multi-hop QA datasets (HotpotQA~\cite{hotpotQA} and 2Wiki~\cite{2wiki}) and encourages search actions as part of long-chain reasoning.
\ole

Implementations of both models are available within the PyTerrier-RAG experimental framework~\cite{pyterrier_rag,pyterrier}.
At each retrieval iteration, the context consists of the top-3 documents retrieved by E5~\cite{E5} from a 2018 Wikipedia snapshot, following the corpus setting in the FlashRAG benchmark~\cite{flashRAG}.

\begin{table}[t]
\centering
\caption{
Trajectory statistics on the development sets for Search-R1 and R1-Searcher.
F1 denotes final answer quality; $\Delta$F1 denotes trajectory-level utility compared with the initial probed answer before retrieval; $\pazocal{U}$ denotes average per-iteration utility; $n$ is average trajectory length.
}
\label{table:dataset_summary}
\begin{adjustbox}{width=0.9\columnwidth}
\begin{tabular}{llrrrrr}
\toprule
Pipeline & Dataset & F1 & $\Delta$F1 & $\pazocal{U}$ & $n$ & \#Q \\
\midrule
\multirow{3}{*}{Search-R1}
& HotpotQA & 0.549 & 0.265 & 0.097 & 2.730 & 7405 \\
& 2Wiki    & 0.429 & 0.155 & 0.045 & 3.463 & 12576 \\
& MuSiQue  & 0.274 & 0.155 & 0.047 & 3.337 & 2417 \\
\midrule
\multirow{3}{*}{R1-Searcher}
& HotpotQA & 0.529 & 0.408 & 0.177 & 2.311 & 7405 \\
& 2Wiki    & 0.454 & 0.386 & 0.161 & 2.386 & 12576 \\
& MuSiQue  & 0.272 & 0.223 & 0.081 & 2.718 & 2417 \\
\bottomrule
\end{tabular}
\end{adjustbox}
\end{table}

\para{Datasets} We evaluate on three multi-hop QA datasets with diverse reasoning complexity and question structures.

\ols
\li \textbf{HotpotQA}~\cite{hotpotQA}: 
A widely used multi-hop QA benchmark where questions are constructed around pairs of Wikipedia entities.
It mainly represents relatively shallow two-hop reasoning.
\li \textbf{2Wiki}~\cite{2wiki}: A multi-hop QA benchmark with more diverse reasoning structures than HotpotQA.
\li \textbf{MuSiQue}~\cite{Musique}: A multi-hop QA benchmark requiring various reasoning hops and reasoning graph topologies, providing cases for studying more complex trajectories.
\ole

We evaluate partial answer quality and utility using F1 score against the ground truth answers~\cite{evaluatingQAEval}, and train supervised predictors and sequential prediction heads on the combined training splits of the three datasets before evaluating them on their development splits.
\footnote{The source code is available at \url{https://github.com/DanielTian97/agentic_rag_predictions}.}
Table~\ref{table:dataset_summary} summarises the trajectory statistics of the evaluated agentic RAG pipelines on each dataset.

\section{Results}\label{s:results}

In this section, we answer the three research questions.
Section~\ref{ss:for_rq1} analyses the evolution of partial answer quality and partial utility in agentic RAG trajectories.
Section~\ref{ss:for_rq2} evaluates whether trajectory-derived signals can predict these intermediate answer states.
Section~\ref{ss:for_rq3} further tests whether such predictions can guide early stopping to improve efficiency.

\begin{figure}[t]
    \centering

    \begin{subfigure}{0.48\columnwidth}
        \centering
        \includegraphics[width=\linewidth]{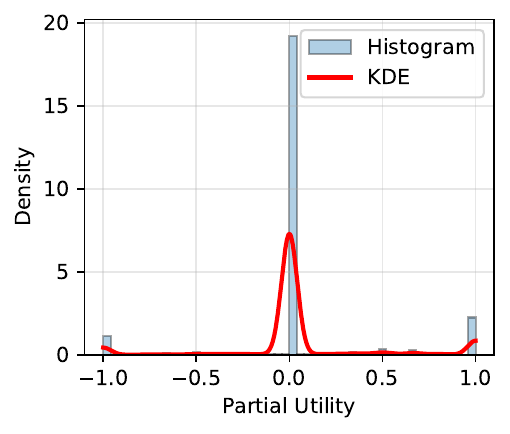}
        \caption{Search-R1}
        \label{fig:left_plot}
    \end{subfigure}
    \hfill
    \begin{subfigure}{0.48\columnwidth}
        \centering
        \includegraphics[width=\linewidth]{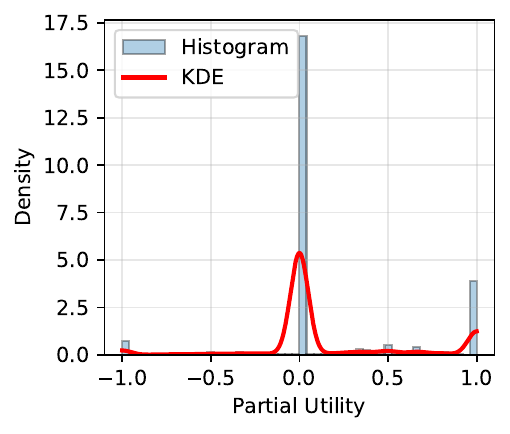}
        \caption{R1-Searcher}
        \label{fig:right_plot}
    \end{subfigure}

    \caption{Distribution of partial utility $\pazocal{U}_i$, instantiated as the change in F1 score, across retrieval-reasoning iterations in Search-R1 (left) and R1-Searcher (right). Results are aggregated over the three evaluated datasets. Histograms show the empirical distribution, and KDE curves provide a smoothed view. Both distributions are sharply concentrated around zero. Positive and negative tails correspond to iterations that improve or degrade the probed answer, respectively.}
    \label{fig:distribution}
\end{figure}

\begin{figure}[tb]
    \centering

    \begin{subfigure}{0.48\columnwidth}
        \centering
        \includegraphics[width=\linewidth]{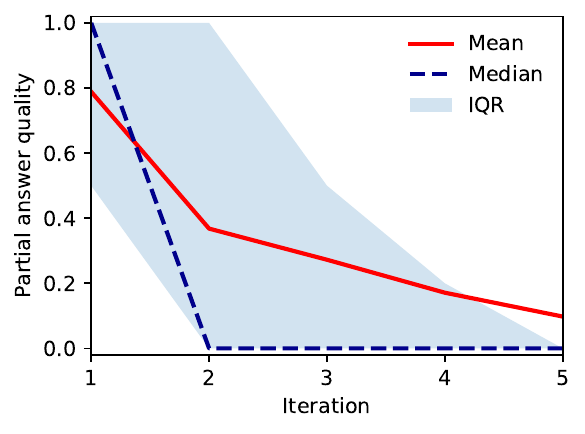}
        \caption{Negative Utility (Search-R1)}
        \label{fig:rq1_r1_negative_utility}
    \end{subfigure}
    \hfill
    \begin{subfigure}{0.48\columnwidth}
        \centering
        \includegraphics[width=\linewidth]{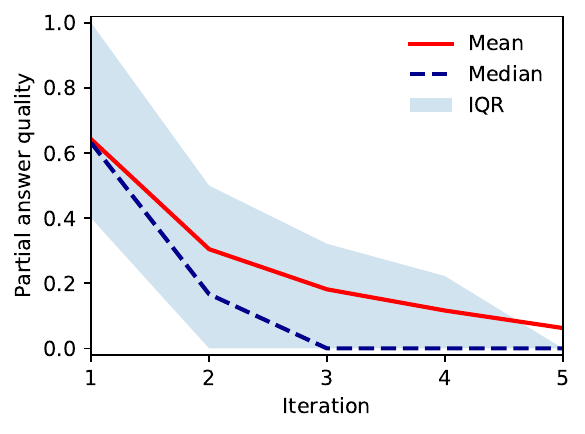}
        \caption{Negative Utility (R1-Searcher)}
        \label{fig:rq1_r1searcher_negative_utility}
    \end{subfigure}

    \vspace{0.4em}

    \begin{subfigure}{0.48\columnwidth}
        \centering
        \includegraphics[width=\linewidth]{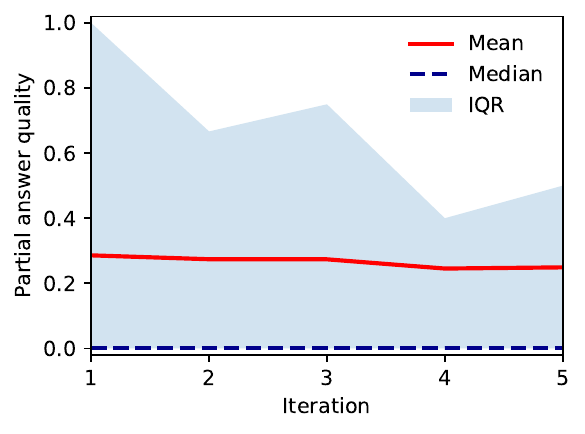}
        \caption{Zero Utility (Search-R1)}
        \label{fig:rq1_r1_zero_utility}
    \end{subfigure}
    \hfill
    \begin{subfigure}{0.48\columnwidth}
        \centering
        \includegraphics[width=\linewidth]{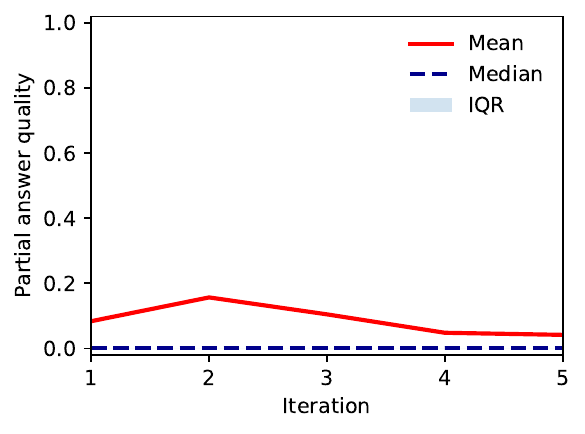}
        \caption{Zero Utility (R1-Searcher)}
        \label{fig:rq1_r1searcher_zero_utility}
    \end{subfigure}

    \vspace{0.4em}

    \begin{subfigure}{0.48\columnwidth}
        \centering
        \includegraphics[width=\linewidth]{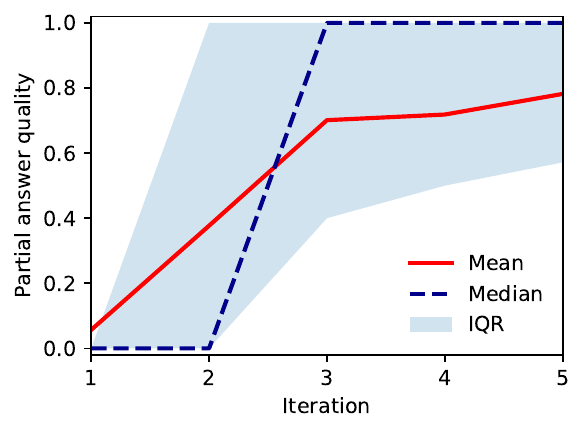}
        \caption{Positive Utility (Search-R1)}
        \label{fig:rq1_r1_positive_utility}
    \end{subfigure}
    \hfill
    \begin{subfigure}{0.48\columnwidth}
        \centering
        \includegraphics[width=\linewidth]{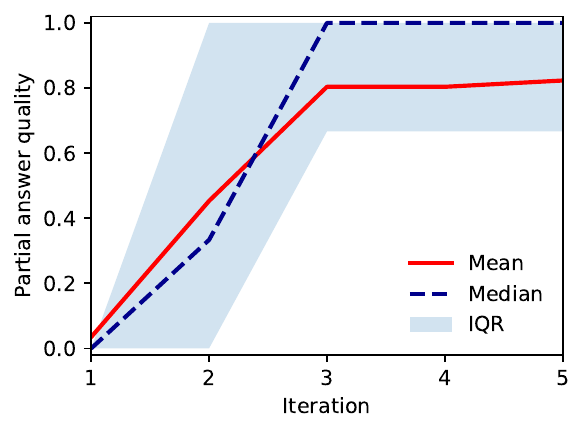}
        \caption{Positive Utility (R1-Searcher)}
        \label{fig:rq1_r1searcher_positive_utility}
    \end{subfigure}

    \caption{Aggregate partial answer quality patterns grouped by trajectory-level net utility across the three evaluated datasets.
    Rows correspond to negative-, zero-, and positive-utility trajectories, while columns correspond to Search-R1 and R1-Searcher.
    Solid and dashed lines show the mean and median, respectively; shaded regions denote the interquartile range (IQR), indicating variability in partial answer quality across trajectories at each iteration.}
    Statistics are computed over trajectories that reach each iteration.
    \label{fig:quality_traces}
\end{figure}

\subsection{RQ-1: Decisive Iterations are Sparsely Distributed in Agentic RAG}\label{ss:for_rq1}

We first analyse how intermediate answer states evolve along agentic RAG trajectories.
Figure~\ref{fig:distribution} shows the distribution of partial utility $\pazocal{U}_i$ across retrieval-reasoning iterations on the merged development sets of the three datasets: HotpotQA, 2Wiki, and MuSiQue.
For both Search-R1 and R1-Searcher, the distribution is sharply concentrated around zero, indicating that most iterations do not substantially change partial answer quality.
This suggests that partial answer quality is not steadily improved at every step in agentic RAG.
Instead, the majority of iterations contribute zero utility, while only a smaller portion of iterations cause clear positive or negative changes in the probed answer quality.

The two extremes of the distributions further indicate that retrieval reasoning iterations can be either beneficial or harmful.
Although these non-zero-utility iterations are relatively sparse, they play a decisive role in determining the trajectory-level net utility, i.e., the overall change between the initial probed answer before retrieval and the final answer after the agentic RAG process.
This motivates a closer analysis of trajectory shapes: if the accumulated utility of a trajectory is driven by a small number of decisive iterations, then trajectories with positive, zero, and negative net utility may exhibit different patterns of answer-quality evolution.

Figure~\ref{fig:quality_traces} shows aggregate partial-answer-quality patterns after grouping trajectories by their net utility, using questions from the development splits of the three evaluated datasets. The summary statistics reveal distinct behaviours across groups.
Positive-utility trajectories generally improve rapidly in the early iterations and then plateau; in later iterations, partial answer quality tends to remain saturated or slightly decrease.
For negative-utility trajectories, partial answer quality often drops sharply in early iterations, and later iterations usually fail to recover the lost quality or even further degrade the answer.
Zero-utility trajectories may fluctuate, but show no net improvement by the end.

To conclude RQ-1, these results obtained by in-trajectory probing show that \textit{decisive iterations are sparse in agentic RAG trajectories and often occur early}.
Most later iterations have limited measured utility, either preserving a plateaued answer quality or failing to recover from earlier degradation.
This suggests \textit{a mismatch between the model's stopping behaviour and its intermediate answer quality}, motivating RQ-2, where we study whether partial answer quality and partial utility can be predicted during the agentic RAG process.

\begin{table*}[t]
\centering
\caption{Prediction effectiveness for partial utility and partial answer quality.
Rows compare different combinations of unsupervised features (Unsp.), supervised predictors (Sup.), and probing-based confidence signals (Prob.). The upper rows show the results in Search-R1, while the lower rows show the results in R1-Searcher.
Results are reported for sequential windows $w \in \{1,3,5\}$ using Pearson's $r$ and Kendall's $\tau$.
The best result for predicting each metric in each column is boldfaced, and the best results for each agentic RAG pipeline are further underlined. A $\dagger$ indicates a significant improvement over $w=1$ ($p<0.05$ by Fisher's $z$ test).}\label{table:main_table}
\begin{adjustbox}{width=0.9\textwidth}
\begin{tabular}{c ccc SS SS SS SS SS SS r}
\toprule
&  &  &  & \multicolumn{6}{c}{Partial Utility Prediction} & \multicolumn{6}{c}{Partial Answer Quality Prediction} \\
\cmidrule(lr){5-10} \cmidrule(lr){11-16}
&  &  &  & \multicolumn{2}{c}{$w$=1} & \multicolumn{2}{c}{$w$=3} & \multicolumn{2}{c}{$w$=5} & \multicolumn{2}{c}{$w$=1} & \multicolumn{2}{c}{$w$=3} & \multicolumn{2}{c}{$w$=5}\\
\cmidrule(lr){5-6} \cmidrule(lr){7-8} \cmidrule(lr){9-10} \cmidrule(lr){11-12} \cmidrule(lr){13-14} \cmidrule(lr){15-16}
& Unsup. & Sup. &  Prob. & $r$ & $\tau$ & $r$ & $\tau$ & $r$ & $\tau$ & $r$ & $\tau$ & $r$ & $\tau$ & $r$ & $\tau$ & {\scriptsize Row ID} \\
\cmidrule(lr){1-4} \cmidrule(lr){5-10} \cmidrule(lr){11-16}
\multirow{7}{*}{\rotatebox[origin=c]{90}{Search-R1}} & \checkmark &  &  & 0.201 & 0.140 & 0.216$^\dagger$ & 0.148$^\dagger$ & 0.227$^\dagger$ & 0.162$^\dagger$ & 0.343 & 0.257 & 0.363$^\dagger$ & 0.272$^\dagger$ & 0.358$^\dagger$ & 0.270$^\dagger$ & $^1$\\
&  & \checkmark &  & 0.295 & 0.200 & 0.293 & 0.196 & 0.293 & 0.195 & 0.418 & 0.331 & 0.420 & 0.332 & 0.419 & 0.330 & $^2$ \\
&  &  & \checkmark & 0.166 & 0.109 & 0.184$^\dagger$ & 0.112 & 0.194$^\dagger$ & 0.125$^\dagger$ & 0.345 & 0.257 & 0.352 & 0.263 & 0.351 & 0.264$^\dagger$ & $^3$ \\
& \checkmark & \checkmark &  & 0.299 & 0.200 & 0.299 & 0.199 & 0.294 & 0.194 & 0.422 & 0.332 & 0.424 & 0.335 & 0.424 & 0.335 & $^4$ \\
& \checkmark &  & \checkmark & 0.232 & 0.147 & 0.250$^\dagger$ & 0.162$^\dagger$ & 0.259$^\dagger$ & 0.176$^\dagger$ & 0.379 & 0.285 & 0.387 & 0.293$^\dagger$ & 0.390$^\dagger$ & 0.295$^\dagger$ & $^5$ \\
&  & \checkmark & \checkmark & 0.306 & \textbf{0.206} & 0.302 & 0.200 & 0.306 & 0.204 & 0.431 & 0.341 & 0.432 & 0.341 & 0.433 & 0.341 & $^6$ \\
& \checkmark & \checkmark &  \checkmark & \textbf{0.308} & 0.204 & \underline{\textbf{0.312}} & \underline{\textbf{0.208}} & \textbf{0.311} & \textbf{0.205} & \underline{\textbf{0.438}} & \underline{\textbf{0.345}} & \textbf{0.434} & \textbf{0.344} & \textbf{0.436} & \textbf{0.344} & $^7$ \\
\cmidrule{1-16}
\multirow{7}{*}{\rotatebox[origin=c]{90}{R1-Searcher}} & \checkmark &  &  & 0.200 & 0.154 & 0.232 & 0.189$^\dagger$ & 0.220$^\dagger$ & 0.181$^\dagger$ & 0.222 & 0.170 & 0.302$^\dagger$ & 0.221$^\dagger$ & 0.307$^\dagger$ & 0.227$^\dagger$ & $^8$ \\
&  & \checkmark &  & 0.309 & 0.243 & 0.313 & 0.245 & 0.310 & 0.242 & 0.390 & 0.314 & 0.391 & 0.314 & 0.389 & 0.310 & $^9$ \\
&  &  & \checkmark & 0.093 & 0.063 & 0.108$^\dagger$ & 0.089$^\dagger$ & 0.062 & 0.035 & 0.090 & 0.048 & 0.095 & 0.055 & 0.107$^\dagger$ & 0.070$^\dagger$ & $^{10}$ \\
& \checkmark & \checkmark &  & 0.314 & \textbf{0.247} & 0.315 & \textbf{0.248} & \textbf{0.321} & \underline{\textbf{0.250}} & \textbf{0.393} & \textbf{0.317} & 0.398 & 0.319 & \textbf{0.398} & \underline{\textbf{0.320}} & $^{11}$ \\
& \checkmark &  & \checkmark & 0.229 & 0.178 & 0.234 & 0.186$^\dagger$ & 0.232$^\dagger$ & 0.187$^\dagger$ & 0.257 & 0.191 & 0.293$^\dagger$ & 0.215$^\dagger$ & 0.293$^\dagger$ & 0.216$^\dagger$ & $^{12}$ \\
&  & \checkmark & \checkmark & 0.309 & 0.242 & 0.313 & 0.244 & 0.311 & 0.244 & 0.379 & 0.308 & 0.396$^\dagger$ & 0.316$^\dagger$ & 0.393$^\dagger$ & 0.315$^\dagger$ & $^{13}$ \\
& \checkmark & \checkmark & \checkmark & \textbf{0.312} & 0.244 & \textbf{0.320} & 0.247 & \underline{\textbf{0.321}} & 0.249 & 0.388 & 0.313 & \underline{\textbf{0.401}}$^\dagger$ & \underline{\textbf{0.320}}$^\dagger$ & 0.394 & 0.317 & $^{14}$ \\
\bottomrule
\end{tabular}

\end{adjustbox}
\label{table:main_table}
\end{table*}

\begin{table}[t]
\centering
\small
\caption{Ablation of unsupervised signal sources for partial answer quality prediction on Search-R1 with window sizes $w \in \{1,3,5\}$.
The best result in each column is bold-faced; $\dagger$ indicates that all combined significantly improves over the best individual signal source under the same window size ($p < 0.05$ by Fisher's $z$ test).}
\label{table:unsup_ablation}
\begin{adjustbox}{width=\columnwidth}
\begin{tabular}{l ll ll ll}
\toprule
\multirow{2}{*}{Signals} 
& \multicolumn{2}{c}{$w=1$}
& \multicolumn{2}{c}{$w=3$}
& \multicolumn{2}{c}{$w=5$} \\
\cmidrule(lr){2-3}\cmidrule(lr){4-5}\cmidrule(lr){6-7}
& $r$ & $\tau$ & $r$ & $\tau$ & $r$ & $\tau$ \\
\midrule
Intra-iteration      & .323 & .239 & .351 & .260 & .347 & .260 \\
Inter-iteration      & .312 & .233 & .331 & .249 & .334 & .251 \\
Question-iteration  & .306 & .223 & .326 & .239 & .330 & .247 \\
All combined         & \textbf{.343}$^\dagger$ & \textbf{.257}$^\dagger$ & \textbf{.363}$^\dagger$ & \textbf{.272}$^\dagger$ & \textbf{.358}$^\dagger$ & \textbf{.270}$^\dagger$ \\
\bottomrule
\end{tabular}
\end{adjustbox}
\end{table}

\subsection{RQ-2: Intermediate Answer State Can Be Predicted from Trajectory-Derived Signals}\label{ss:for_rq2}

We next examine how effectively the prediction methods proposed in Section~\ref{ss:scheme} estimate partial answer quality and partial utility.
Table~\ref{table:main_table} reports the prediction performance obtained by combining different groups of predictors under varying context window sizes (the number of recent iterations included in sequential prediction).
The three input groups correspond to unsupervised features, supervised predictors, and probing-based confidence signals.
The final estimates are produced by an MLP prediction head over sequential windows $w \in \{1,3,5\}$.

First, we observe from Table~\ref{table:main_table} that partial answer quality is consistently more predictable than partial utility.
Across both Search-R1 and R1-Searcher, the correlations for partial answer quality are higher than those for partial utility in almost all experimental settings.
This is expected because partial utility is a \emph{difference} signal: it depends not only on the quality of the current retrieval-reasoning iteration, but also on the previous answer state and the extent to which newly retrieved evidence changes the model's intermediate answer.
These factors are harder to infer from externally observable trajectory signals than the quality of a single probed answer.
In Search-R1, the best partial answer quality prediction reaches Pearson's $r=0.438$ and Kendall's $\tau=0.345$, while the best partial utility prediction reaches $r=0.321$ and $\tau=0.208$.
A similar pattern is observed for R1-Searcher.
Thus, although both targets are predictable, partial answer quality provides a more reliable target for estimating intermediate answer states without ground-truth evaluation.

The second observation from Table~\ref{table:main_table} is that the three signal groups provide complementary information.
Using only unsupervised features already yields positive correlations, showing that diagnostic signals such as retrieval quality and inter-iteration changes contain useful evidence about partial answer quality and partial utility.
Among the single signal groups, the supervised predictors are the strongest in both pipelines (rows 2 and 9), reflecting the benefit of direct optimisation for the proposed targets.
Combining unsupervised features with supervised predictors only slightly improves prediction effectiveness, but it can make the prediction more interpretable by linking the estimate to retrieval quality, trajectory change, and query-iteration alignment.
Probing-based confidence is also informative, although its contribution is weaker, especially for R1-Searcher.
This suggests that generation confidence alone is not always a reliable indicator of intermediate answer quality in agentic RAG trajectories, which involve long reasoning generations.
Overall, the all-combined setting achieves the best or near-best results in both pipelines (rows 7 and 14), showing that the three signal groups provide complementary evidence.

The third observation from Table~\ref{table:main_table} is that larger windows help most when unsupervised signals are included, as seen in rows 1, 4, 5, 7 and their R1-Searcher counterparts.
Larger windows provide the prediction head with a longer history of unsupervised signal values, allowing it to compare the current iteration against recent trajectory patterns.
In contrast, supervised predictors benefit less consistently from larger windows, suggesting that they already encode much of the relevant information centred on the current iteration.
Thus, sequential aggregation is especially useful for weaker or more indirect signals, where recent trajectory history can compensate for limited local evidence.

Beyond the main prediction results, Table~\ref{table:unsup_ablation} ablates the unsupervised signal sources for partial answer quality prediction.
Since these unsupervised signals are diagnostic indicators, this ablation helps identify which types of trajectory relations are more informative for estimating answer quality.
Among individual sources, intra-iteration signals are generally strongest, suggesting that current retrieval quality and local iteration evidence are useful indicators of the intermediate answer state.
Combining all three sources consistently improves over each individual source, showing that retrieval quality, inter-iteration change, and question--iteration alignment provide complementary diagnostic evidence.
The gain from increasing the prediction window from $w=1$ to $w=3$ suggests that unsupervised predictors benefit from a short history of recent iterations, while the limited gain at $w=5$ indicates that longer histories add little additional information.

To conclude RQ-2, we find that \textit{both partial answer quality and partial utility can be predicted from trajectory-derived signals, but partial answer quality is the more reliable prediction target.}
Combining unsupervised diagnostics with supervised predictors yields strong results, while probing-based confidence provides additional but optional evidence.
Sequential prediction further helps when historical unsupervised signals are included.

\begin{figure}[t]
    \centering
    \includegraphics[width=\linewidth]{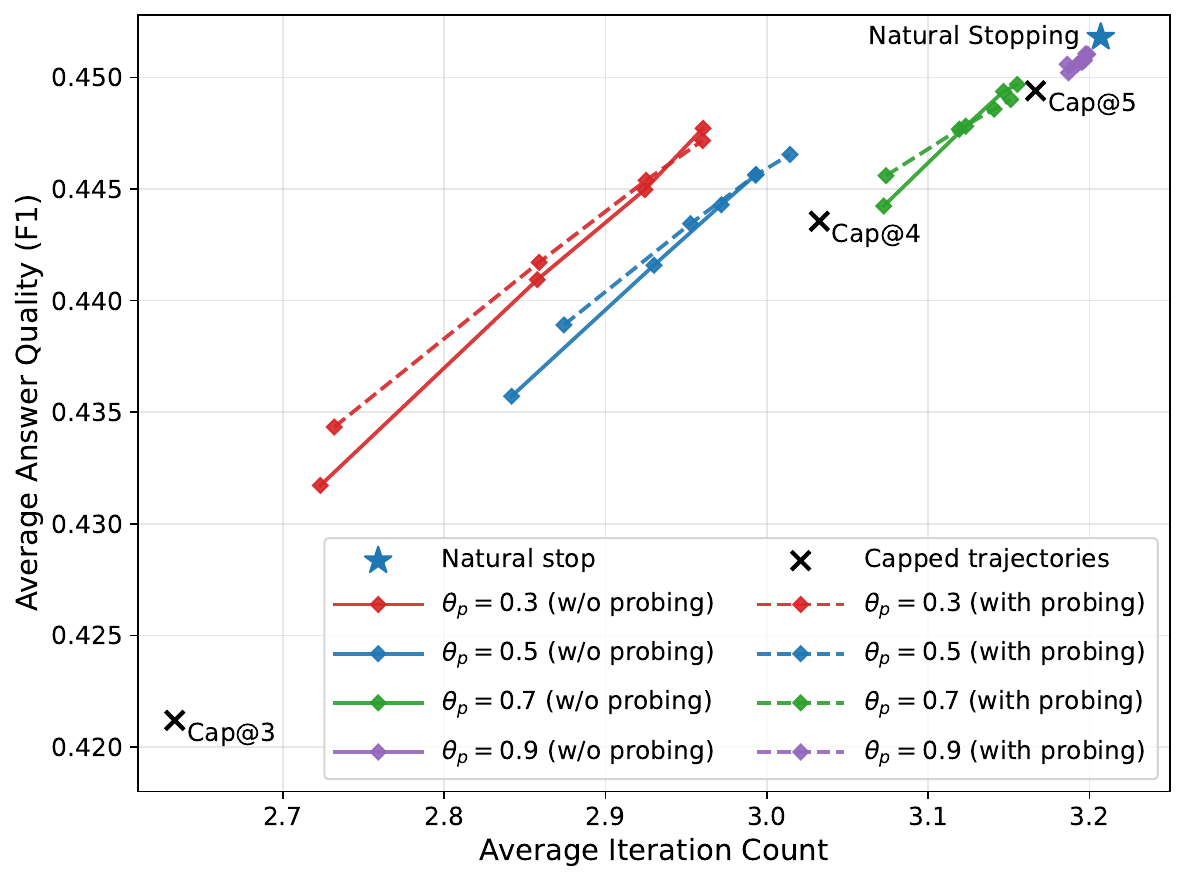}
    \caption{Efficiency-effectiveness trade-off of the proposed stopping controller on Search-R1. The star marks natural stopping, black crosses mark fixed iteration caps, and coloured curves show prediction-guided stopping under different $\theta_P$ values. Markers vary $\theta_U$ from 0.1 to 0.4 from left to right; solid and dashed lines denote stopping without and with probing.
}\label{fig:efficiency_effectiveness_tradeoff}
\end{figure}

\subsection{RQ-3: Prediction-Guided Stopping Improves Efficiency}\label{ss:for_rq3}
\label{ss:for_rq3}

We finally examine whether intermediate answer state predictions can be used to control the agentic RAG process and improve its efficiency.
Following the state-based early stopping method in Section~\ref{s:early_stopping}, we evaluate the controller on the Search-R1 pipeline using Unsup.+Sup. as the main predictor combination.
We also test Unsup.+Sup.+Prob., which slightly improves prediction performance in Table~\ref{table:main_table} at the cost of additional answer-probing overhead.
We evaluate success by the efficiency--effectiveness trade-off: reducing the average iteration count while preserving final answer quality.

Figure~\ref{fig:efficiency_effectiveness_tradeoff} compares our prediction-guided stopping heuristic with natural stopping and the fixed-cap baselines, Cap@$k$, defined in Section~\ref{ss:implementations}.
Natural stopping gives the highest answer quality at the highest iteration cost, while fixed caps save iterations through uniform truncation but cause a sharp quality drop.
In contrast, our prediction-guided heuristic provides a smoother trade-off curve: lowering the quality threshold $\theta_P$ and utility threshold $\theta_U$ makes the stopping policy more aggressive, reducing iterations at the cost of a higher risk of answer-quality loss.

The main advantage of prediction-guided stopping is that it makes trace-specific decisions.
Unlike fixed caps, which stop all trajectories at the same length, our heuristic uses predicted partial answer quality and partial utility to identify whether a trajectory has reached a useful answer state or entered a low-utility region.
As shown in Figure~\ref{fig:efficiency_effectiveness_tradeoff}, with $\theta_P=0.3$, setting $\theta_U=0.2$ reduces the average iteration count to 2.86, 10.89\% fewer than natural stopping's 3.21, while preserving 97.60\% of its answer quality. Increasing $\theta_U$ to 0.3 preserves 98.49\% of answer quality with only 0.06 additional iterations on average compared with $\theta_U=0.2$.
This indicates that intermediate answer state predictions can reduce redundant iterations with limited quality loss, while the thresholds $\theta_P$ and $\theta_U$ provide control over the efficiency-effectiveness trade-off.

We also compare stopping variants with and without probing.
When probing is enabled, each checked iteration requires an additional answer generation conditioned on the trajectory prefix, so its overhead can be comparable to a normal generation step.
In contrast, the non-probing predictors use signals already produced by the agentic RAG trajectory: unsupervised features require only lightweight computations such as similarity, overlap, or retrieval score statistics, and supervised predictors use compact SBERT-based regressors rather than the LLM backbone.
As shown by the dashed and solid curves in Figure~\ref{fig:efficiency_effectiveness_tradeoff}, adding probing brings only limited improvement in answer quality.
This suggests that, in practical settings, prediction-guided stopping can often be applied without probing, avoiding its additional generation cost.

To conclude RQ-3, we find that \textit{prediction-guided stopping reduces average iterations while preserving most of the answer quality achieved by natural stopping}.
This shows that partial answer quality and partial utility predictions can serve as practical control signals for improving the efficiency-effectiveness trade-off in agentic RAG.

\section{Future Work}\label{s:conclusion}

In this paper, we studied intermediate answer states in agentic RAG.
We introduced an in-trajectory probing framework that forces the model to produce an answer after each retrieval-reasoning iteration, allowing us to measure partial answer quality and partial utility along the trajectory.
Our analysis shows that decisive iterations are sparse: many iterations contribute little measured utility, while meaningful changes in answer quality often occur early.
This reveals a mismatch between natural stopping behaviour and the point at which answer quality has already saturated or degraded.

We further formulated two prediction tasks: partial answer quality prediction and partial utility prediction.
Experiments show that trajectory-derived signals are informative for both tasks, but partial answer quality is more reliably predicted than partial utility.
Supervised predictors provide strong estimates of intermediate answer states, while unsupervised diagnostics and probing-based confidence signals offer complementary information.
Using these predictions for early stopping, a simple prediction-guided controller reduces the average number of retrieval-reasoning iterations while preserving most of the answer quality achieved by natural stopping.

Our study focuses on prediction rather than optimised controller design.
The early-stopping heuristic is intentionally simple, and adaptive policies may further improve the efficiency-effectiveness trade-off.
Moreover, partial answer quality is measured using QA-style metrics such as F1, which may miss finer-grained changes in intermediate answers.
Future work could extend intermediate answer state prediction to advanced agentic RAG pipelines and complex tasks, such as open-ended long-form generation.

\section*{GenAI Usage Disclosure}
No generative AI was used to produce the research ideas, results, or future plans described in this paper. Generative AI was used for limited language polishing and proofreading.

\bibliographystyle{ACM-Reference-Format}

\begin{thebibliography}{67}


\ifx \showCODEN    \undefined \def \showCODEN     #1{\unskip}     \fi
\ifx \showISBNx    \undefined \def \showISBNx     #1{\unskip}     \fi
\ifx \showISBNxiii \undefined \def \showISBNxiii  #1{\unskip}     \fi
\ifx \showISSN     \undefined \def \showISSN      #1{\unskip}     \fi
\ifx \showLCCN     \undefined \def \showLCCN      #1{\unskip}     \fi
\ifx \shownote     \undefined \def \shownote      #1{#1}          \fi
\ifx \showarticletitle \undefined \def \showarticletitle #1{#1}   \fi
\ifx \showURL      \undefined \def \showURL       {\relax}        \fi
\providecommand\bibfield[2]{#2}
\providecommand\bibinfo[2]{#2}
\providecommand\natexlab[1]{#1}
\providecommand\showeprint[2][]{arXiv:#2}
\makeatletter
\@ifundefined{NAT@parse@date}{}{\let\NAT@parse@date@orig\NAT@parse@date}
\@ifundefined{NAT@parse@date}{}{\def\NAT@parse@date#1#2#3#4#5#6@@{\NAT@parse@date@orig#1#2#3#4#5#6@@\def\NAT@tempyear{0000}\def\NAT@tempexlab{{?}}\ifx\NAT@year\NAT@tempyear\ifx\NAT@exlab\NAT@tempexlab\def\NAT@date{[n.\,d.]}\else\edef\NAT@date{[n.\,d.]\NAT@exlab}\fi\fi}}
\makeatother

\bibitem[Aksitov et~al\mbox{.}(2023)]%
        {bamboogle}
\bibfield{author}{\bibinfo{person}{Renat Aksitov}, \bibinfo{person}{Sobhan Miryoosefi}, \bibinfo{person}{Zonglin Li}, \bibinfo{person}{Daliang Li}, \bibinfo{person}{Sheila Babayan}, \bibinfo{person}{Kavya Kopparapu}, \bibinfo{person}{Zachary Fisher}, \bibinfo{person}{Ruiqi Guo}, \bibinfo{person}{Sushant Prakash}, \bibinfo{person}{Pranesh Srinivasan}, \bibinfo{person}{Manzil Zaheer}, \bibinfo{person}{Felix Yu}, {and} \bibinfo{person}{Sanjiv Kumar}.} \bibinfo{year}{2023}\natexlab{}.
\newblock \bibinfo{title}{ReST meets ReAct: Self-Improvement for Multi-Step Reasoning LLM Agent}.
\newblock
\showeprint[arxiv]{2312.10003}~[cs.CL]
\urldef\tempurl%
\url{https://arxiv.org/abs/2312.10003}
\showURL{%
\tempurl}


\bibitem[Amiraz et~al\mbox{.}(2025)]%
        {understandingIrrelevantContextInRAG}
\bibfield{author}{\bibinfo{person}{Chen Amiraz}, \bibinfo{person}{Florin Cuconasu}, \bibinfo{person}{Simone Filice}, {and} \bibinfo{person}{Zohar Karnin}.} \bibinfo{year}{2025}\natexlab{}.
\newblock \showarticletitle{The Distracting Effect: Understanding Irrelevant Passages in {RAG}}. In \bibinfo{booktitle}{\emph{Proceedings of the 63rd Annual Meeting of the Association for Computational Linguistics (Volume 1: Long Papers)}}. \bibinfo{publisher}{Association for Computational Linguistics}, \bibinfo{pages}{18228--18258}.
\newblock
\showISBNx{979-8-89176-251-0}
\urldef\tempurl%
\url{https://aclanthology.org/2025.acl-long.892/}
\showURL{%
\tempurl}


\bibitem[Arabzadeh et~al\mbox{.}(2021)]%
        {bertQPP}
\bibfield{author}{\bibinfo{person}{Negar Arabzadeh}, \bibinfo{person}{Maryam Khodabakhsh}, {and} \bibinfo{person}{Ebrahim Bagheri}.} \bibinfo{year}{2021}\natexlab{}.
\newblock \showarticletitle{BERT-QPP: Contextualized Pre-trained transformers for Query Performance Prediction}. In \bibinfo{booktitle}{\emph{Proceedings of the 30th ACM International Conference on Information \& Knowledge Management}} \emph{(\bibinfo{series}{CIKM '21})}. \bibinfo{publisher}{Association for Computing Machinery}, \bibinfo{pages}{2857–2861}.
\newblock
\showISBNx{9781450384469}
\urldef\tempurl%
\url{https://doi.org/10.1145/3459637.3482063}
\showURL{%
\tempurl}


\bibitem[Chen et~al\mbox{.}(2019)]%
        {evaluatingQAEval}
\bibfield{author}{\bibinfo{person}{Anthony Chen}, \bibinfo{person}{Gabriel Stanovsky}, \bibinfo{person}{Sameer Singh}, {and} \bibinfo{person}{Matt Gardner}.} \bibinfo{year}{2019}\natexlab{}.
\newblock \showarticletitle{Evaluating Question Answering Evaluation}. In \bibinfo{booktitle}{\emph{Proceedings of the 2nd Workshop on Machine Reading for Question Answering}}. \bibinfo{publisher}{Association for Computational Linguistics}, \bibinfo{address}{Hong Kong, China}, \bibinfo{pages}{119--124}.
\newblock
\urldef\tempurl%
\url{https://aclanthology.org/D19-5817/}
\showURL{%
\tempurl}


\bibitem[Chen et~al\mbox{.}(2024)]%
        {internalStateForHallucinationDetection}
\bibfield{author}{\bibinfo{person}{Chao Chen}, \bibinfo{person}{Kai Liu}, \bibinfo{person}{Ze Chen}, \bibinfo{person}{Yi Gu}, \bibinfo{person}{Yue Wu}, \bibinfo{person}{Mingyuan Tao}, \bibinfo{person}{Zhihang Fu}, {and} \bibinfo{person}{Jieping Ye}.} \bibinfo{year}{2024}\natexlab{}.
\newblock \showarticletitle{{INSIDE}: {LLM}s' Internal States Retain the Power of Hallucination Detection}. In \bibinfo{booktitle}{\emph{The Twelfth International Conference on Learning Representations}}.
\newblock
\urldef\tempurl%
\url{https://openreview.net/forum?id=Zj12nzlQbz}
\showURL{%
\tempurl}


\bibitem[Dado et~al\mbox{.}(2026)]%
        {RAGPredictionForQA}
\bibfield{author}{\bibinfo{person}{Or Dado}, \bibinfo{person}{David Carmel}, {and} \bibinfo{person}{Oren Kurland}.} \bibinfo{year}{2026}\natexlab{}.
\newblock \bibinfo{title}{Rag Performance Prediction for Question Answering}.
\newblock
\showeprint[arxiv]{2604.07985}~[cs.CL]
\urldef\tempurl%
\url{https://arxiv.org/abs/2604.07985}
\showURL{%
\tempurl}


\bibitem[Dassen et~al\mbox{.}(2026)]%
        {FACTUM}
\bibfield{author}{\bibinfo{person}{Maxime Dassen}, \bibinfo{person}{Rebecca Kotula}, \bibinfo{person}{Kenton Murray}, \bibinfo{person}{Andrew Yates}, \bibinfo{person}{Dawn Lawrie}, \bibinfo{person}{Efsun Kayi}, \bibinfo{person}{James Mayfield}, {and} \bibinfo{person}{Kevin Duh}.} \bibinfo{year}{2026}\natexlab{}.
\newblock \showarticletitle{FACTUM: Mechanistic Detection of Citation Hallucination in Long-Form RAG}. In \bibinfo{booktitle}{\emph{Advances in Information Retrieval: 48th European Conference on Information Retrieval, ECIR 2026, Delft, The Netherlands, March 29 – April 2, 2026, Proceedings, Part I}}. \bibinfo{publisher}{Springer-Verlag}, \bibinfo{pages}{272–288}.
\newblock
\showISBNx{978-3-032-21288-7}
\urldef\tempurl%
\url{https://doi.org/10.1007/978-3-032-21289-4_18}
\showURL{%
\tempurl}


\bibitem[Datta et~al\mbox{.}(2022a)]%
        {deepQPP}
\bibfield{author}{\bibinfo{person}{Suchana Datta}, \bibinfo{person}{Debasis Ganguly}, \bibinfo{person}{Derek Greene}, {and} \bibinfo{person}{Mandar Mitra}.} \bibinfo{year}{2022}\natexlab{a}.
\newblock \showarticletitle{Deep-QPP: A Pairwise Interaction-based Deep Learning Model for Supervised Query Performance Prediction}. In \bibinfo{booktitle}{\emph{Proceedings of the Fifteenth ACM International Conference on Web Search and Data Mining}} \emph{(\bibinfo{series}{WSDM '22})}. \bibinfo{publisher}{Association for Computing Machinery}, \bibinfo{pages}{201–209}.
\newblock
\showISBNx{9781450391320}
\urldef\tempurl%
\url{https://doi.org/10.1145/3488560.3498491}
\showURL{%
\tempurl}


\bibitem[Datta et~al\mbox{.}(2022b)]%
        {WRIG}
\bibfield{author}{\bibinfo{person}{Suchana Datta}, \bibinfo{person}{Debasis Ganguly}, \bibinfo{person}{Mandar Mitra}, {and} \bibinfo{person}{Derek Greene}.} \bibinfo{year}{2022}\natexlab{b}.
\newblock \showarticletitle{A Relative Information Gain-based Query Performance Prediction Framework with Generated Query Variants}.
\newblock \bibinfo{journal}{\emph{ACM Trans. Inf. Syst.}} \bibinfo{volume}{41}, \bibinfo{number}{2}, Article \bibinfo{articleno}{38} (\bibinfo{date}{dec} \bibinfo{year}{2022}), \bibinfo{numpages}{31}~pages.
\newblock
\showISSN{1046-8188}
\href{https://doi.org/10.1145/3545112}{doi:\nolinkurl{10.1145/3545112}}


\bibitem[Ebrahimi et~al\mbox{.}(2024)]%
        {ContextRichQPP}
\bibfield{author}{\bibinfo{person}{Sajad Ebrahimi}, \bibinfo{person}{Maryam Khodabakhsh}, \bibinfo{person}{Negar Arabzadeh}, {and} \bibinfo{person}{Ebrahim Bagheri}.} \bibinfo{year}{2024}\natexlab{}.
\newblock \showarticletitle{Estimating Query Performance Through Rich Contextualized Query Representations}. In \bibinfo{booktitle}{\emph{Advances in Information Retrieval}}. \bibinfo{publisher}{Springer Nature Switzerland}, \bibinfo{pages}{49--58}.
\newblock
\showISBNx{978-3-031-56066-8}


\bibitem[Faggioli et~al\mbox{.}(2023)]%
        {denseQPP}
\bibfield{author}{\bibinfo{person}{Guglielmo Faggioli}, \bibinfo{person}{Nicola Ferro}, \bibinfo{person}{Cristina~Ioana Muntean}, \bibinfo{person}{Raffaele Perego}, {and} \bibinfo{person}{Nicola Tonellotto}.} \bibinfo{year}{2023}\natexlab{}.
\newblock \showarticletitle{A Geometric Framework for Query Performance Prediction in Conversational Search}. In \bibinfo{booktitle}{\emph{Proceedings of the 46th International ACM SIGIR Conference on Research and Development in Information Retrieval}} \emph{(\bibinfo{series}{SIGIR '23})}. \bibinfo{publisher}{Association for Computing Machinery}, \bibinfo{pages}{1355–1365}.
\newblock
\showISBNx{9781450394086}
\urldef\tempurl%
\url{https://doi.org/10.1145/3539618.3591625}
\showURL{%
\tempurl}


\bibitem[Fang et~al\mbox{.}(2024)]%
        {trace}
\bibfield{author}{\bibinfo{person}{Jinyuan Fang}, \bibinfo{person}{Zaiqiao Meng}, {and} \bibinfo{person}{Craig MacDonald}.} \bibinfo{year}{2024}\natexlab{}.
\newblock \showarticletitle{{TRACE} the Evidence: Constructing Knowledge-Grounded Reasoning Chains for Retrieval-Augmented Generation}. In \bibinfo{booktitle}{\emph{Findings of the Association for Computational Linguistics: EMNLP 2024}}. \bibinfo{pages}{8472--8494}.
\newblock
\href{https://doi.org/10.18653/v1/2024.findings-emnlp.496}{doi:\nolinkurl{10.18653/v1/2024.findings-emnlp.496}}


\bibitem[Fang et~al\mbox{.}(2025)]%
        {kiRAG}
\bibfield{author}{\bibinfo{person}{Jinyuan Fang}, \bibinfo{person}{Zaiqiao Meng}, {and} \bibinfo{person}{Craig MacDonald}.} \bibinfo{year}{2025}\natexlab{}.
\newblock \showarticletitle{{K}i{RAG}: Knowledge-Driven Iterative Retriever for Enhancing Retrieval-Augmented Generation}. In \bibinfo{booktitle}{\emph{Proceedings of the 63rd Annual Meeting of the Association for Computational Linguistics (Volume 1: Long Papers)}}. \bibinfo{publisher}{Association for Computational Linguistics}, \bibinfo{pages}{18969--18985}.
\newblock
\showISBNx{979-8-89176-251-0}
\href{https://doi.org/10.18653/v1/2025.acl-long.929}{doi:\nolinkurl{10.18653/v1/2025.acl-long.929}}


\bibitem[Farquhar et~al\mbox{.}(2024)]%
        {detectHallucinations}
\bibfield{author}{\bibinfo{person}{Sebastian Farquhar}, \bibinfo{person}{Jannik Kossen}, \bibinfo{person}{Lorenz Kuhn}, {and} \bibinfo{person}{Yarin Gal}.} \bibinfo{year}{2024}\natexlab{}.
\newblock \showarticletitle{Detecting hallucinations in large language models using semantic entropy}.
\newblock \bibinfo{journal}{\emph{Nature (London)}} \bibinfo{volume}{630}, \bibinfo{number}{8017} (\bibinfo{year}{2024}), \bibinfo{pages}{625--630}.
\newblock
\showISBNx{0028-0836;1476-4687;}


\bibitem[Gao et~al\mbox{.}(2024)]%
        {ragReview}
\bibfield{author}{\bibinfo{person}{Yunfan Gao}, \bibinfo{person}{Yun Xiong}, \bibinfo{person}{Xinyu Gao}, \bibinfo{person}{Kangxiang Jia}, \bibinfo{person}{Jinliu Pan}, \bibinfo{person}{Yuxi Bi}, \bibinfo{person}{Yi Dai}, \bibinfo{person}{Jiawei Sun}, \bibinfo{person}{Meng Wang}, {and} \bibinfo{person}{Haofen Wang}.} \bibinfo{year}{2024}\natexlab{}.
\newblock \bibinfo{title}{Retrieval-Augmented Generation for Large Language Models: A Survey}.
\newblock
\showeprint[arxiv]{2312.10997}~[cs.CL]
\urldef\tempurl%
\url{https://arxiv.org/abs/2312.10997}
\showURL{%
\tempurl}


\bibitem[Ho et~al\mbox{.}(2020)]%
        {2wiki}
\bibfield{author}{\bibinfo{person}{Xanh Ho}, \bibinfo{person}{Anh-Khoa~Duong Nguyen}, \bibinfo{person}{Saku Sugawara}, {and} \bibinfo{person}{Akiko Aizawa}.} \bibinfo{year}{2020}\natexlab{}.
\newblock \bibinfo{title}{Constructing A Multi-hop QA Dataset for Comprehensive Evaluation of Reasoning Steps}.
\newblock
\showeprint[arxiv]{2011.01060}~[cs.CL]
\urldef\tempurl%
\url{https://arxiv.org/abs/2011.01060}
\showURL{%
\tempurl}


\bibitem[Ji et~al\mbox{.}(2024)]%
        {llmInternalStateRevealHallucinationRisk}
\bibfield{author}{\bibinfo{person}{Ziwei Ji}, \bibinfo{person}{Delong Chen}, \bibinfo{person}{Etsuko Ishii}, \bibinfo{person}{Samuel Cahyawijaya}, \bibinfo{person}{Yejin Bang}, \bibinfo{person}{Bryan Wilie}, {and} \bibinfo{person}{Pascale Fung}.} \bibinfo{year}{2024}\natexlab{}.
\newblock \showarticletitle{{LLM} Internal States Reveal Hallucination Risk Faced With a Query}. In \bibinfo{booktitle}{\emph{Proceedings of the 7th BlackboxNLP Workshop: Analyzing and Interpreting Neural Networks for NLP}}. \bibinfo{publisher}{Association for Computational Linguistics}, \bibinfo{address}{Miami, Florida, US}, \bibinfo{pages}{88--104}.
\newblock
\href{https://doi.org/10.18653/v1/2024.blackboxnlp-1.6}{doi:\nolinkurl{10.18653/v1/2024.blackboxnlp-1.6}}


\bibitem[Ji et~al\mbox{.}(2023)]%
        {surveyOfHallucinations}
\bibfield{author}{\bibinfo{person}{Ziwei Ji}, \bibinfo{person}{Nayeon Lee}, \bibinfo{person}{Rita Frieske}, \bibinfo{person}{Tiezheng Yu}, \bibinfo{person}{Dan Su}, \bibinfo{person}{Yan Xu}, \bibinfo{person}{Etsuko Ishii}, \bibinfo{person}{Ye~Jin Bang}, \bibinfo{person}{Andrea Madotto}, {and} \bibinfo{person}{Pascale Fung}.} \bibinfo{year}{2023}\natexlab{}.
\newblock \showarticletitle{Survey of Hallucination in Natural Language Generation}.
\newblock \bibinfo{journal}{\emph{ACM Comput. Surv.}} \bibinfo{volume}{55}, \bibinfo{number}{12}, Article \bibinfo{articleno}{248} (\bibinfo{date}{March} \bibinfo{year}{2023}), \bibinfo{numpages}{38}~pages.
\newblock
\showISSN{0360-0300}
\href{https://doi.org/10.1145/3571730}{doi:\nolinkurl{10.1145/3571730}}


\bibitem[Jiang et~al\mbox{.}(2025)]%
        {tcRAG}
\bibfield{author}{\bibinfo{person}{Xinke Jiang}, \bibinfo{person}{Yue Fang}, \bibinfo{person}{Rihong Qiu}, \bibinfo{person}{Haoyu Zhang}, \bibinfo{person}{Yongxin Xu}, \bibinfo{person}{Hao Chen}, \bibinfo{person}{Wentao Zhang}, \bibinfo{person}{Ruizhe Zhang}, \bibinfo{person}{Yuchen Fang}, \bibinfo{person}{Xinyu Ma}, \bibinfo{person}{Xu Chu}, \bibinfo{person}{Junfeng Zhao}, {and} \bibinfo{person}{Yasha Wang}.} \bibinfo{year}{2025}\natexlab{}.
\newblock \showarticletitle{{TC}{--}{RAG}: {T}uring{--}Complete {RAG}{'}s Case study on Medical {LLM} Systems}. In \bibinfo{booktitle}{\emph{Proceedings of the 63rd Annual Meeting of the Association for Computational Linguistics (Volume 1: Long Papers)}}. \bibinfo{publisher}{Association for Computational Linguistics}, \bibinfo{pages}{11400--11426}.
\newblock
\showISBNx{979-8-89176-251-0}
\href{https://doi.org/10.18653/v1/2025.acl-long.558}{doi:\nolinkurl{10.18653/v1/2025.acl-long.558}}


\bibitem[Jiang et~al\mbox{.}(2023)]%
        {FLARE}
\bibfield{author}{\bibinfo{person}{Zhengbao Jiang}, \bibinfo{person}{Frank Xu}, \bibinfo{person}{Luyu Gao}, \bibinfo{person}{Zhiqing Sun}, \bibinfo{person}{Qian Liu}, \bibinfo{person}{Jane Dwivedi-Yu}, \bibinfo{person}{Yiming Yang}, \bibinfo{person}{Jamie Callan}, {and} \bibinfo{person}{Graham Neubig}.} \bibinfo{year}{2023}\natexlab{}.
\newblock \showarticletitle{Active Retrieval Augmented Generation}. In \bibinfo{booktitle}{\emph{Proceedings of the 2023 Conference on Empirical Methods in Natural Language Processing}}. \bibinfo{publisher}{Association for Computational Linguistics}, \bibinfo{pages}{7969--7992}.
\newblock
\href{https://doi.org/10.18653/v1/2023.emnlp-main.495}{doi:\nolinkurl{10.18653/v1/2023.emnlp-main.495}}


\bibitem[Jin et~al\mbox{.}(2025a)]%
        {searchR1}
\bibfield{author}{\bibinfo{person}{Bowen Jin}, \bibinfo{person}{Hansi Zeng}, \bibinfo{person}{Zhenrui Yue}, \bibinfo{person}{Jinsung Yoon}, \bibinfo{person}{Sercan Arik}, \bibinfo{person}{Dong Wang}, \bibinfo{person}{Hamed Zamani}, {and} \bibinfo{person}{Jiawei Han}.} \bibinfo{year}{2025}\natexlab{a}.
\newblock \bibinfo{title}{Search-R1: Training LLMs to Reason and Leverage Search Engines with Reinforcement Learning}.
\newblock
\showeprint[arxiv]{2503.09516}~[cs.CL]
\urldef\tempurl%
\url{https://arxiv.org/abs/2503.09516}
\showURL{%
\tempurl}


\bibitem[Jin et~al\mbox{.}(2025b)]%
        {flashRAG}
\bibfield{author}{\bibinfo{person}{Jiajie Jin}, \bibinfo{person}{Yutao Zhu}, \bibinfo{person}{Zhicheng Dou}, \bibinfo{person}{Guanting Dong}, \bibinfo{person}{Xinyu Yang}, \bibinfo{person}{Chenghao Zhang}, \bibinfo{person}{Tong Zhao}, \bibinfo{person}{Zhao Yang}, {and} \bibinfo{person}{Ji-Rong Wen}.} \bibinfo{year}{2025}\natexlab{b}.
\newblock \showarticletitle{FlashRAG: A Modular Toolkit for Efficient Retrieval-Augmented Generation Research}. In \bibinfo{booktitle}{\emph{Companion Proceedings of the ACM on Web Conference 2025}} \emph{(\bibinfo{series}{WWW '25})}. \bibinfo{publisher}{Association for Computing Machinery}, \bibinfo{pages}{737–740}.
\newblock
\showISBNx{9798400713316}
\urldef\tempurl%
\url{https://doi.org/10.1145/3701716.3715313}
\showURL{%
\tempurl}


\bibitem[Joshi et~al\mbox{.}(2017)]%
        {triviaQA}
\bibfield{author}{\bibinfo{person}{Mandar Joshi}, \bibinfo{person}{Eunsol Choi}, \bibinfo{person}{Daniel Weld}, {and} \bibinfo{person}{Luke Zettlemoyer}.} \bibinfo{year}{2017}\natexlab{}.
\newblock \showarticletitle{{T}rivia{QA}: A Large Scale Distantly Supervised Challenge Dataset for Reading Comprehension}. In \bibinfo{booktitle}{\emph{Proceedings of the 55th Annual Meeting of the Association for Computational Linguistics (Volume 1: Long Papers)}}. \bibinfo{publisher}{Association for Computational Linguistics}, \bibinfo{pages}{1601--1611}.
\newblock
\urldef\tempurl%
\url{https://aclanthology.org/P17-1147/}
\showURL{%
\tempurl}


\bibitem[Kadavath et~al\mbox{.}(2022)]%
        {llmKnowsWhatItKnows}
\bibfield{author}{\bibinfo{person}{Saurav Kadavath}, \bibinfo{person}{Tom Conerly}, \bibinfo{person}{Amanda Askell}, \bibinfo{person}{Tom Henighan}, \bibinfo{person}{Dawn Drain}, \bibinfo{person}{Ethan Perez}, \bibinfo{person}{Nicholas Schiefer}, \bibinfo{person}{Zac Hatfield-Dodds}, \bibinfo{person}{Nova DasSarma}, \bibinfo{person}{Eli Tran-Johnson}, \bibinfo{person}{Scott Johnston}, \bibinfo{person}{Sheer El-Showk}, \bibinfo{person}{Andy Jones}, \bibinfo{person}{Nelson Elhage}, \bibinfo{person}{Tristan Hume}, \bibinfo{person}{Anna Chen}, \bibinfo{person}{Yuntao Bai}, \bibinfo{person}{Sam Bowman}, \bibinfo{person}{Stanislav Fort}, \bibinfo{person}{Deep Ganguli}, \bibinfo{person}{Danny Hernandez}, \bibinfo{person}{Josh Jacobson}, \bibinfo{person}{Jackson Kernion}, \bibinfo{person}{Shauna Kravec}, \bibinfo{person}{Liane Lovitt}, \bibinfo{person}{Kamal Ndousse}, \bibinfo{person}{Catherine Olsson}, \bibinfo{person}{Sam Ringer}, \bibinfo{person}{Dario Amodei}, \bibinfo{person}{Tom Brown}, \bibinfo{person}{Jack
  Clark}, \bibinfo{person}{Nicholas Joseph}, \bibinfo{person}{Ben Mann}, \bibinfo{person}{Sam McCandlish}, \bibinfo{person}{Chris Olah}, {and} \bibinfo{person}{Jared Kaplan}.} \bibinfo{year}{2022}\natexlab{}.
\newblock \bibinfo{title}{Language Models (Mostly) Know What They Know}.
\newblock
\showeprint[arxiv]{2207.05221}~[cs.CL]
\urldef\tempurl%
\url{https://arxiv.org/abs/2207.05221}
\showURL{%
\tempurl}


\bibitem[Kwiatkowski et~al\mbox{.}(2019)]%
        {NQ}
\bibfield{author}{\bibinfo{person}{Tom Kwiatkowski}, \bibinfo{person}{Jennimaria Palomaki}, \bibinfo{person}{Olivia Redfield}, \bibinfo{person}{Michael Collins}, \bibinfo{person}{Ankur Parikh}, \bibinfo{person}{Chris Alberti}, \bibinfo{person}{Danielle Epstein}, \bibinfo{person}{Illia Polosukhin}, \bibinfo{person}{Jacob Devlin}, \bibinfo{person}{Kenton Lee}, \bibinfo{person}{Kristina Toutanova}, \bibinfo{person}{Llion Jones}, \bibinfo{person}{Matthew Kelcey}, \bibinfo{person}{Ming-Wei Chang}, \bibinfo{person}{Andrew~M. Dai}, \bibinfo{person}{Jakob Uszkoreit}, \bibinfo{person}{Quoc Le}, {and} \bibinfo{person}{Slav Petrov}.} \bibinfo{year}{2019}\natexlab{}.
\newblock \showarticletitle{Natural Questions: A Benchmark for Question Answering Research}.
\newblock \bibinfo{journal}{\emph{Transactions of the Association for Computational Linguistics}}  \bibinfo{volume}{7} (\bibinfo{year}{2019}), \bibinfo{pages}{452--466}.
\newblock
\urldef\tempurl%
\url{https://aclanthology.org/Q19-1026/}
\showURL{%
\tempurl}


\bibitem[Lewis et~al\mbox{.}(2020)]%
        {ragInKnowledgeIntensiveNLP}
\bibfield{author}{\bibinfo{person}{Patrick Lewis}, \bibinfo{person}{Ethan Perez}, \bibinfo{person}{Aleksandra Piktus}, \bibinfo{person}{Fabio Petroni}, \bibinfo{person}{Vladimir Karpukhin}, \bibinfo{person}{Naman Goyal}, \bibinfo{person}{Heinrich K\"{u}ttler}, \bibinfo{person}{Mike Lewis}, \bibinfo{person}{Wen-tau Yih}, \bibinfo{person}{Tim Rockt\"{a}schel}, \bibinfo{person}{Sebastian Riedel}, {and} \bibinfo{person}{Douwe Kiela}.} \bibinfo{year}{2020}\natexlab{}.
\newblock \showarticletitle{Retrieval-augmented generation for knowledge-intensive {NLP} tasks}. In \bibinfo{booktitle}{\emph{Proceedings of the 34th International Conference on Neural Information Processing Systems}} \emph{(\bibinfo{series}{NIPS '20})}. \bibinfo{publisher}{Curran Associates Inc.}, Article \bibinfo{articleno}{793}, \bibinfo{numpages}{16}~pages.
\newblock
\showISBNx{9781713829546}
\urldef\tempurl%
\url{https://arxiv.org/abs/2005.11401}
\showURL{%
\tempurl}


\bibitem[Li et~al\mbox{.}(2025b)]%
        {MeCo}
\bibfield{author}{\bibinfo{person}{Wenjun Li}, \bibinfo{person}{Dexun Li}, \bibinfo{person}{Kuicai Dong}, \bibinfo{person}{Cong Zhang}, \bibinfo{person}{Hao Zhang}, \bibinfo{person}{Weiwen Liu}, \bibinfo{person}{Yasheng Wang}, \bibinfo{person}{Ruiming Tang}, {and} \bibinfo{person}{Yong Liu}.} \bibinfo{year}{2025}\natexlab{b}.
\newblock \showarticletitle{Adaptive Tool Use in Large Language Models with Meta-Cognition Trigger}. In \bibinfo{booktitle}{\emph{Proceedings of the 63rd Annual Meeting of the Association for Computational Linguistics (Volume 1: Long Papers)}}. \bibinfo{publisher}{Association for Computational Linguistics}, \bibinfo{address}{Vienna, Austria}.
\newblock
\showISBNx{979-8-89176-251-0}
\href{https://doi.org/10.18653/v1/2025.acl-long.655}{doi:\nolinkurl{10.18653/v1/2025.acl-long.655}}


\bibitem[Li et~al\mbox{.}(2025a)]%
        {searchO1}
\bibfield{author}{\bibinfo{person}{Xiaoxi Li}, \bibinfo{person}{Guanting Dong}, \bibinfo{person}{Jiajie Jin}, \bibinfo{person}{Yuyao Zhang}, \bibinfo{person}{Yujia Zhou}, \bibinfo{person}{Yutao Zhu}, \bibinfo{person}{Peitian Zhang}, {and} \bibinfo{person}{Zhicheng Dou}.} \bibinfo{year}{2025}\natexlab{a}.
\newblock \bibinfo{title}{Search-o1: Agentic Search-Enhanced Large Reasoning Models}.
\newblock
\showeprint[arxiv]{2501.05366}~[cs.AI]
\urldef\tempurl%
\url{https://arxiv.org/abs/2501.05366}
\showURL{%
\tempurl}


\bibitem[Li et~al\mbox{.}(2025c)]%
        {towardsAgenticRAGwithDeepReasoning}
\bibfield{author}{\bibinfo{person}{Yangning Li}, \bibinfo{person}{Weizhi Zhang}, \bibinfo{person}{Yuyao Yang}, \bibinfo{person}{Wei-Chieh Huang}, \bibinfo{person}{Yaozu Wu}, \bibinfo{person}{Junyu Luo}, \bibinfo{person}{Yuanchen Bei}, \bibinfo{person}{Henry~Peng Zou}, \bibinfo{person}{Xiao Luo}, \bibinfo{person}{Yusheng Zhao}, \bibinfo{person}{Chunkit Chan}, \bibinfo{person}{Yankai Chen}, \bibinfo{person}{Zhongfen Deng}, \bibinfo{person}{Yinghui Li}, \bibinfo{person}{Hai-Tao Zheng}, \bibinfo{person}{Dongyuan Li}, \bibinfo{person}{Renhe Jiang}, \bibinfo{person}{Ming Zhang}, \bibinfo{person}{Yangqiu Song}, {and} \bibinfo{person}{Philip~S. Yu}.} \bibinfo{year}{2025}\natexlab{c}.
\newblock \showarticletitle{A Survey of {RAG}-Reasoning Systems in Large Language Models}. In \bibinfo{booktitle}{\emph{Findings of the Association for Computational Linguistics: EMNLP 2025}}. \bibinfo{publisher}{Association for Computational Linguistics}, \bibinfo{pages}{12120--12145}.
\newblock
\showISBNx{979-8-89176-335-7}
\href{https://doi.org/10.18653/v1/2025.findings-emnlp.648}{doi:\nolinkurl{10.18653/v1/2025.findings-emnlp.648}}


\bibitem[Liu et~al\mbox{.}(2024)]%
        {lostInTheMiddle}
\bibfield{author}{\bibinfo{person}{Nelson~F. Liu}, \bibinfo{person}{Kevin Lin}, \bibinfo{person}{John Hewitt}, \bibinfo{person}{Ashwin Paranjape}, \bibinfo{person}{Michele Bevilacqua}, \bibinfo{person}{Fabio Petroni}, {and} \bibinfo{person}{Percy Liang}.} \bibinfo{year}{2024}\natexlab{}.
\newblock \showarticletitle{Lost in the Middle: How Language Models Use Long Contexts}.
\newblock \bibinfo{journal}{\emph{Transactions of the Association for Computational Linguistics}}  \bibinfo{volume}{12} (\bibinfo{year}{2024}), \bibinfo{pages}{157--173}.
\newblock
\href{https://doi.org/10.1162/tacl_a_00638}{doi:\nolinkurl{10.1162/tacl_a_00638}}


\bibitem[Macdonald et~al\mbox{.}(2025)]%
        {pyterrier_rag}
\bibfield{author}{\bibinfo{person}{Craig Macdonald}, \bibinfo{person}{Jinyuan Fang}, \bibinfo{person}{Andrew Parry}, {and} \bibinfo{person}{Zaiqiao Meng}.} \bibinfo{year}{2025}\natexlab{}.
\newblock \showarticletitle{Constructing and Evaluating Declarative RAG Pipelines in PyTerrier}. In \bibinfo{booktitle}{\emph{Proceedings of the 48th International ACM SIGIR Conference on Research and Development in Information Retrieval}} \emph{(\bibinfo{series}{SIGIR '25})}. \bibinfo{publisher}{Association for Computing Machinery}, \bibinfo{pages}{4035–4040}.
\newblock
\showISBNx{9798400715921}
\urldef\tempurl%
\url{https://doi.org/10.1145/3726302.3730150}
\showURL{%
\tempurl}


\bibitem[Macdonald et~al\mbox{.}(2021)]%
        {pyterrier}
\bibfield{author}{\bibinfo{person}{Craig Macdonald}, \bibinfo{person}{Nicola Tonellotto}, \bibinfo{person}{Sean MacAvaney}, {and} \bibinfo{person}{Iadh Ounis}.} \bibinfo{year}{2021}\natexlab{}.
\newblock \showarticletitle{{PyTerrier}: Declarative Experimentation in Python from BM25 to Dense Retrieval}. In \bibinfo{booktitle}{\emph{Proceedings of the 30th ACM International Conference on Information \& Knowledge Management}} \emph{(\bibinfo{series}{CIKM '21})}. \bibinfo{publisher}{Association for Computing Machinery}, \bibinfo{address}{New York, NY, USA}, \bibinfo{pages}{4526–4533}.
\newblock
\showISBNx{9781450384469}
\href{https://doi.org/10.1145/3459637.3482013}{doi:\nolinkurl{10.1145/3459637.3482013}}


\bibitem[Marjanovic et~al\mbox{.}(2024)]%
        {conflictsInRAG}
\bibfield{author}{\bibinfo{person}{Sara~Vera Marjanovic}, \bibinfo{person}{Haeun Yu}, \bibinfo{person}{Pepa Atanasova}, \bibinfo{person}{Maria Maistro}, \bibinfo{person}{Christina Lioma}, {and} \bibinfo{person}{Isabelle Augenstein}.} \bibinfo{year}{2024}\natexlab{}.
\newblock \showarticletitle{{DYNAMICQA}: Tracing Internal Knowledge Conflicts in Language Models}. In \bibinfo{booktitle}{\emph{Findings of the Association for Computational Linguistics: EMNLP 2024}}. \bibinfo{publisher}{Association for Computational Linguistics}, \bibinfo{address}{Miami, Florida, USA}, \bibinfo{pages}{14346--14360}.
\newblock
\urldef\tempurl%
\url{https://aclanthology.org/2024.findings-emnlp.838/}
\showURL{%
\tempurl}


\bibitem[Mo et~al\mbox{.}(2025)]%
        {surveyOnConversationalSearch}
\bibfield{author}{\bibinfo{person}{Fengran Mo}, \bibinfo{person}{Kelong Mao}, \bibinfo{person}{Ziliang Zhao}, \bibinfo{person}{Hongjin Qian}, \bibinfo{person}{Haonan Chen}, \bibinfo{person}{Yiruo Cheng}, \bibinfo{person}{Xiaoxi Li}, \bibinfo{person}{Yutao Zhu}, \bibinfo{person}{Zhicheng Dou}, {and} \bibinfo{person}{Jian-Yun Nie}.} \bibinfo{year}{2025}\natexlab{}.
\newblock \showarticletitle{A Survey of Conversational Search}.
\newblock \bibinfo{journal}{\emph{ACM Trans. Inf. Syst.}} \bibinfo{volume}{43}, \bibinfo{number}{6}, Article \bibinfo{articleno}{167} (\bibinfo{date}{Sept.} \bibinfo{year}{2025}), \bibinfo{numpages}{50}~pages.
\newblock
\showISSN{1046-8188}
\urldef\tempurl%
\url{https://doi.org/10.1145/3759453}
\showURL{%
\tempurl}


\bibitem[Park et~al\mbox{.}(2025)]%
        {stopRAG}
\bibfield{author}{\bibinfo{person}{Jaewan Park}, \bibinfo{person}{Solbee Cho}, {and} \bibinfo{person}{Jay-Yoon Lee}.} \bibinfo{year}{2025}\natexlab{}.
\newblock \showarticletitle{Stop-{RAG}: Value-Based Retrieval Control for Iterative {RAG}}. In \bibinfo{booktitle}{\emph{First Workshop on Multi-Turn Interactions in Large Language Models}}.
\newblock
\urldef\tempurl%
\url{https://openreview.net/forum?id=fN4PJNYgs3}
\showURL{%
\tempurl}


\bibitem[Qin et~al\mbox{.}(2025)]%
        {AMBER}
\bibfield{author}{\bibinfo{person}{Qitao Qin}, \bibinfo{person}{Yucong Luo}, \bibinfo{person}{Yihang Lu}, \bibinfo{person}{Zhibo Chu}, \bibinfo{person}{Xiaoman Liu}, {and} \bibinfo{person}{Xianwei Meng}.} \bibinfo{year}{2025}\natexlab{}.
\newblock \showarticletitle{Towards Adaptive Memory-Based Optimization for Enhanced Retrieval-Augmented Generation}. In \bibinfo{booktitle}{\emph{Findings of the Association for Computational Linguistics: ACL 2025}}. \bibinfo{publisher}{Association for Computational Linguistics}, \bibinfo{pages}{7991--8004}.
\newblock
\showISBNx{979-8-89176-256-5}
\urldef\tempurl%
\url{https://aclanthology.org/2025.findings-acl.418/}
\showURL{%
\tempurl}


\bibitem[Raiber and Kurland(2014)]%
        {QPP-goal}
\bibfield{author}{\bibinfo{person}{Fiana Raiber} {and} \bibinfo{person}{Oren Kurland}.} \bibinfo{year}{2014}\natexlab{}.
\newblock \showarticletitle{Query-performance prediction: setting the expectations straight}. In \bibinfo{booktitle}{\emph{Proceedings of the 37th International ACM SIGIR Conference on Research \& Development in Information Retrieval}} \emph{(\bibinfo{series}{SIGIR '14})}. \bibinfo{publisher}{Association for Computing Machinery}, \bibinfo{pages}{13–22}.
\newblock
\showISBNx{9781450322577}
\href{https://doi.org/10.1145/2600428.2609581}{doi:\nolinkurl{10.1145/2600428.2609581}}


\bibitem[Reimers and Gurevych(2019)]%
        {SBERT}
\bibfield{author}{\bibinfo{person}{Nils Reimers} {and} \bibinfo{person}{Iryna Gurevych}.} \bibinfo{year}{2019}\natexlab{}.
\newblock \showarticletitle{Sentence-BERT: Sentence Embeddings using Siamese BERT-Networks}. In \bibinfo{booktitle}{\emph{Proceedings of the 2019 Conference on Empirical Methods in Natural Language Processing}}. \bibinfo{publisher}{Association for Computational Linguistics}.
\newblock
\urldef\tempurl%
\url{https://arxiv.org/abs/1908.10084}
\showURL{%
\tempurl}


\bibitem[Roitman et~al\mbox{.}(2017)]%
        {RSD}
\bibfield{author}{\bibinfo{person}{Haggai Roitman}, \bibinfo{person}{Shai Erera}, {and} \bibinfo{person}{Bar Weiner}.} \bibinfo{year}{2017}\natexlab{}.
\newblock \showarticletitle{Robust Standard Deviation Estimation for Query Performance Prediction}. In \bibinfo{booktitle}{\emph{Proceedings of the ACM SIGIR International Conference on Theory of Information Retrieval}} \emph{(\bibinfo{series}{ICTIR '17})}. \bibinfo{publisher}{Association for Computing Machinery}, \bibinfo{pages}{245–248}.
\newblock
\showISBNx{9781450344906}
\href{https://doi.org/10.1145/3121050.3121087}{doi:\nolinkurl{10.1145/3121050.3121087}}


\bibitem[Roy et~al\mbox{.}(2024)]%
        {selfRAG}
\bibfield{author}{\bibinfo{person}{Nirmal Roy}, \bibinfo{person}{Leonardo F.~R. Ribeiro}, \bibinfo{person}{Rexhina Blloshmi}, {and} \bibinfo{person}{Kevin Small}.} \bibinfo{year}{2024}\natexlab{}.
\newblock \showarticletitle{Learning When to Retrieve, What to Rewrite, and How to Respond in Conversational {QA}}. In \bibinfo{booktitle}{\emph{Findings of the Association for Computational Linguistics: EMNLP 2024}}. \bibinfo{publisher}{Association for Computational Linguistics}, \bibinfo{pages}{10604--10625}.
\newblock
\urldef\tempurl%
\url{https://aclanthology.org/2024.findings-emnlp.622/}
\showURL{%
\tempurl}


\bibitem[Ru et~al\mbox{.}(2024)]%
        {ragChecker}
\bibfield{author}{\bibinfo{person}{Dongyu Ru}, \bibinfo{person}{Lin Qiu}, \bibinfo{person}{Xiangkun Hu}, \bibinfo{person}{Tianhang Zhang}, \bibinfo{person}{Peng Shi}, \bibinfo{person}{Shuaichen Chang}, \bibinfo{person}{Cheng Jiayang}, \bibinfo{person}{Cunxiang Wang}, \bibinfo{person}{Shichao Sun}, \bibinfo{person}{Huanyu Li}, \bibinfo{person}{Zizhao Zhang}, \bibinfo{person}{Binjie Wang}, \bibinfo{person}{Jiarong Jiang}, \bibinfo{person}{Tong He}, \bibinfo{person}{Zhiguo Wang}, \bibinfo{person}{Pengfei Liu}, \bibinfo{person}{Yue Zhang}, {and} \bibinfo{person}{Zheng Zhang}.} \bibinfo{year}{2024}\natexlab{}.
\newblock \showarticletitle{RAGChecker: A Fine-grained Framework for Diagnosing Retrieval-Augmented Generation}. In \bibinfo{booktitle}{\emph{Advances in Neural Information Processing Systems}}, Vol.~\bibinfo{volume}{37}. \bibinfo{publisher}{Curran Associates, Inc.}, \bibinfo{pages}{21999--22027}.
\newblock
\urldef\tempurl%
\url{https://proceedings.neurips.cc/paper_files/paper/2024/file/27245589131d17368cccdfa990cbf16e-Paper-Datasets_and_Benchmarks_Track.pdf}
\showURL{%
\tempurl}


\bibitem[Salemi and Zamani(2024)]%
        {ragRelevanceLabel}
\bibfield{author}{\bibinfo{person}{Alireza Salemi} {and} \bibinfo{person}{Hamed Zamani}.} \bibinfo{year}{2024}\natexlab{}.
\newblock \showarticletitle{Evaluating Retrieval Quality in Retrieval-Augmented Generation}. In \bibinfo{booktitle}{\emph{Proceedings of the 47th International ACM SIGIR Conference on Research and Development in Information Retrieval}} \emph{(\bibinfo{series}{SIGIR '24})}. \bibinfo{pages}{2395–2400}.
\newblock
\showISBNx{9798400704314}
\urldef\tempurl%
\url{https://doi.org/10.1145/3626772.3657957}
\showURL{%
\tempurl}


\bibitem[Shen et~al\mbox{.}(2023)]%
        {llmAreNotHumanLevelEvaluator}
\bibfield{author}{\bibinfo{person}{Chenhui Shen}, \bibinfo{person}{Liying Cheng}, \bibinfo{person}{Xuan-Phi Nguyen}, \bibinfo{person}{Yang You}, {and} \bibinfo{person}{Lidong Bing}.} \bibinfo{year}{2023}\natexlab{}.
\newblock \showarticletitle{Large Language Models are Not Yet Human-Level Evaluators for Abstractive Summarization}. In \bibinfo{booktitle}{\emph{Findings of the Association for Computational Linguistics: EMNLP 2023}}. \bibinfo{publisher}{Association for Computational Linguistics}, \bibinfo{address}{Singapore}, \bibinfo{pages}{4215--4233}.
\newblock
\urldef\tempurl%
\url{https://aclanthology.org/2023.findings-emnlp.278/}
\showURL{%
\tempurl}


\bibitem[Shtok et~al\mbox{.}(2012)]%
        {NQC}
\bibfield{author}{\bibinfo{person}{Anna Shtok}, \bibinfo{person}{Oren Kurland}, \bibinfo{person}{David Carmel}, \bibinfo{person}{Fiana Raiber}, {and} \bibinfo{person}{Gad Markovits}.} \bibinfo{year}{2012}\natexlab{}.
\newblock \showarticletitle{Predicting Query Performance by Query-Drift Estimation}.
\newblock \bibinfo{journal}{\emph{ACM Trans. Inf. Syst.}} \bibinfo{volume}{30}, \bibinfo{number}{2}, Article \bibinfo{articleno}{11} (\bibinfo{date}{May} \bibinfo{year}{2012}), \bibinfo{numpages}{35}~pages.
\newblock
\showISSN{1046-8188}
\href{https://doi.org/10.1145/2180868.2180873}{doi:\nolinkurl{10.1145/2180868.2180873}}


\bibitem[Singh et~al\mbox{.}(2025)]%
        {agenticRAGSurvey}
\bibfield{author}{\bibinfo{person}{Aditi Singh}, \bibinfo{person}{Abul Ehtesham}, \bibinfo{person}{Saket Kumar}, {and} \bibinfo{person}{Tala~Talaei Khoei}.} \bibinfo{year}{2025}\natexlab{}.
\newblock \bibinfo{title}{Agentic Retrieval-Augmented Generation: A Survey on Agentic RAG}.
\newblock
\showeprint[arxiv]{2501.09136}~[cs.AI]
\urldef\tempurl%
\url{https://arxiv.org/abs/2501.09136}
\showURL{%
\tempurl}


\bibitem[Song et~al\mbox{.}(2025)]%
        {r1Searcher}
\bibfield{author}{\bibinfo{person}{Huatong Song}, \bibinfo{person}{Jinhao Jiang}, \bibinfo{person}{Yingqian Min}, \bibinfo{person}{Jie Chen}, \bibinfo{person}{Zhipeng Chen}, \bibinfo{person}{Wayne~Xin Zhao}, \bibinfo{person}{Lei Fang}, {and} \bibinfo{person}{Ji-Rong Wen}.} \bibinfo{year}{2025}\natexlab{}.
\newblock \bibinfo{title}{R1-Searcher: Incentivizing the Search Capability in LLMs via Reinforcement Learning}.
\newblock
\showeprint[arxiv]{2503.05592}~[cs.AI]
\urldef\tempurl%
\url{https://arxiv.org/abs/2503.05592}
\showURL{%
\tempurl}


\bibitem[Sun et~al\mbox{.}(2025)]%
        {ReDeEp}
\bibfield{author}{\bibinfo{person}{ZhongXiang Sun}, \bibinfo{person}{Xiaoxue Zang}, \bibinfo{person}{Kai Zheng}, \bibinfo{person}{Jun Xu}, \bibinfo{person}{Xiao Zhang}, \bibinfo{person}{Weijie Yu}, \bibinfo{person}{Yang Song}, {and} \bibinfo{person}{Han Li}.} \bibinfo{year}{2025}\natexlab{}.
\newblock \showarticletitle{ReDe{EP}: Detecting Hallucination in Retrieval-Augmented Generation via Mechanistic Interpretability}. In \bibinfo{booktitle}{\emph{The Thirteenth International Conference on Learning Representations}}.
\newblock
\urldef\tempurl%
\url{https://openreview.net/forum?id=ztzZDzgfrh}
\showURL{%
\tempurl}


\bibitem[Thakur et~al\mbox{.}(2025)]%
        {evalForTRECRAG2024}
\bibfield{author}{\bibinfo{person}{Nandan Thakur}, \bibinfo{person}{Ronak Pradeep}, \bibinfo{person}{Shivani Upadhyay}, \bibinfo{person}{Daniel Campos}, \bibinfo{person}{Nick Craswell}, {and} \bibinfo{person}{Jimmy Lin}.} \bibinfo{year}{2025}\natexlab{}.
\newblock \bibinfo{title}{Support Evaluation for the TREC 2024 RAG Track: Comparing Human versus LLM Judges}.
\newblock
\showeprint[arxiv]{2504.15205}~[cs.CL]
\urldef\tempurl%
\url{https://arxiv.org/abs/2504.15205}
\showURL{%
\tempurl}


\bibitem[Tian et~al\mbox{.}(2025a)]%
        {agenticRAGandQPP}
\bibfield{author}{\bibinfo{person}{Fangzheng Tian}, \bibinfo{person}{Jinyuan Fang}, \bibinfo{person}{Debasis Ganguly}, \bibinfo{person}{Zaiqiao Meng}, {and} \bibinfo{person}{Craig Macdonald}.} \bibinfo{year}{2025}\natexlab{a}.
\newblock \bibinfo{title}{Am I on the Right Track? What Can Predicted Query Performance Tell Us about the Search Behaviour of Agentic RAG}.
\newblock
\showeprint[arxiv]{2507.10411}~[cs.IR]
\urldef\tempurl%
\url{https://arxiv.org/abs/2507.10411}
\showURL{%
\tempurl}


\bibitem[Tian et~al\mbox{.}(2025b)]%
        {RelevanceAndUtility}
\bibfield{author}{\bibinfo{person}{Fangzheng Tian}, \bibinfo{person}{Debasis Ganguly}, {and} \bibinfo{person}{Craig Macdonald}.} \bibinfo{year}{2025}\natexlab{b}.
\newblock \showarticletitle{Is Relevance Propagated from Retriever to Generator in RAG?}. In \bibinfo{booktitle}{\emph{Advances in Information Retrieval: 47th European Conference on Information Retrieval, ECIR 2025, Lucca, Italy, April 6–10, 2025, Proceedings, Part I}}. \bibinfo{publisher}{Springer-Verlag}, \bibinfo{pages}{32–48}.
\newblock
\showISBNx{978-3-031-88707-9}
\href{https://doi.org/10.1007/978-3-031-88708-6_3}{doi:\nolinkurl{10.1007/978-3-031-88708-6_3}}


\bibitem[Tian et~al\mbox{.}(2026)]%
        {ragPredictions}
\bibfield{author}{\bibinfo{person}{Fangzheng Tian}, \bibinfo{person}{Debasis Ganguly}, {and} \bibinfo{person}{Craig Macdonald}.} \bibinfo{year}{2026}\natexlab{}.
\newblock \showarticletitle{Predicting Retrieval Utility and Answer Quality in Retrieval-Augmented Generation}. In \bibinfo{booktitle}{\emph{Advances in Information Retrieval: 48th European Conference on Information Retrieval, ECIR 2026, Delft, The Netherlands, March 29 – April 2, 2026, Proceedings, Part I}}. \bibinfo{publisher}{Springer-Verlag}, \bibinfo{pages}{368–385}.
\newblock
\showISBNx{978-3-032-21288-7}
\href{https://doi.org/10.1007/978-3-032-21289-4_24}{doi:\nolinkurl{10.1007/978-3-032-21289-4_24}}


\bibitem[Trivedi et~al\mbox{.}(2022)]%
        {Musique}
\bibfield{author}{\bibinfo{person}{Harsh Trivedi}, \bibinfo{person}{Niranjan Balasubramanian}, \bibinfo{person}{Tushar Khot}, {and} \bibinfo{person}{Ashish Sabharwal}.} \bibinfo{year}{2022}\natexlab{}.
\newblock \showarticletitle{MuSiQue: Multihop Questions via Single-hop Question Composition}.
\newblock \bibinfo{journal}{\emph{Trans. Assoc. Comput. Linguistics}}  \bibinfo{volume}{10} (\bibinfo{year}{2022}), \bibinfo{pages}{539--554}.
\newblock
\urldef\tempurl%
\url{https://aclanthology.org/2022.tacl-1.31/}
\showURL{%
\tempurl}


\bibitem[Trivedi et~al\mbox{.}(2023)]%
        {iRCoT}
\bibfield{author}{\bibinfo{person}{Harsh Trivedi}, \bibinfo{person}{Niranjan Balasubramanian}, \bibinfo{person}{Tushar Khot}, {and} \bibinfo{person}{Ashish Sabharwal}.} \bibinfo{year}{2023}\natexlab{}.
\newblock \showarticletitle{Interleaving Retrieval with Chain-of-Thought Reasoning for Knowledge-Intensive Multi-Step Questions}. In \bibinfo{booktitle}{\emph{Proceedings of the 61st Annual Meeting of the Association for Computational Linguistics (Volume 1: Long Papers)}}. \bibinfo{pages}{10014--10037}.
\newblock
\urldef\tempurl%
\url{https://arxiv.org/abs/2212.10509}
\showURL{%
\tempurl}


\bibitem[Upadhyay et~al\mbox{.}(2026)]%
        {TREC-RAG-2025-overview}
\bibfield{author}{\bibinfo{person}{Shivani Upadhyay}, \bibinfo{person}{Nandan Thakur}, \bibinfo{person}{Ronak Pradeep}, \bibinfo{person}{Nick Craswell}, \bibinfo{person}{Daniel Campos}, {and} \bibinfo{person}{Jimmy Lin}.} \bibinfo{year}{2026}\natexlab{}.
\newblock \bibinfo{title}{Overview of the TREC 2025 Retrieval Augmented Generation (RAG) Track}.
\newblock
\showeprint[arxiv]{2603.09891}~[cs.IR]
\urldef\tempurl%
\url{https://arxiv.org/abs/2603.09891}
\showURL{%
\tempurl}


\bibitem[Varshney et~al\mbox{.}(2022)]%
        {efficientRAG}
\bibfield{author}{\bibinfo{person}{Neeraj Varshney}, \bibinfo{person}{Man Luo}, {and} \bibinfo{person}{Chitta Baral}.} \bibinfo{year}{2022}\natexlab{}.
\newblock \bibinfo{title}{Can Open-Domain QA Reader Utilize External Knowledge Efficiently like Humans?}
\newblock
\showeprint[arxiv]{2211.12707}~[cs.CL]
\urldef\tempurl%
\url{https://arxiv.org/abs/2211.12707}
\showURL{%
\tempurl}


\bibitem[Vlachou(2024)]%
        {PredRetrFailInConvRec}
\bibfield{author}{\bibinfo{person}{Maria Vlachou}.} \bibinfo{year}{2024}\natexlab{}.
\newblock \emph{\bibinfo{title}{Predicting Retrieval Failures in Conversational Recommendation Systems}}.
\newblock PhD thesis. \bibinfo{school}{University of Glasgow}.
\newblock
\urldef\tempurl%
\url{https://theses.gla.ac.uk/85203/}
\showURL{%
\tempurl}


\bibitem[Vlachou and Macdonald(2024)]%
        {aPairRatio}
\bibfield{author}{\bibinfo{person}{Maria Vlachou} {and} \bibinfo{person}{Craig Macdonald}.} \bibinfo{year}{2024}\natexlab{}.
\newblock \showarticletitle{Coherence-based Query Performance Measures for Dense Retrieval}. In \bibinfo{booktitle}{\emph{Proceedings of the 2024 ACM SIGIR International Conference on Theory of Information Retrieval}} \emph{(\bibinfo{series}{ICTIR '24})}. \bibinfo{publisher}{Association for Computing Machinery}, \bibinfo{pages}{15–24}.
\newblock
\showISBNx{9798400706813}
\urldef\tempurl%
\url{https://doi.org/10.1145/3664190.3672518}
\showURL{%
\tempurl}


\bibitem[Wang et~al\mbox{.}(2026)]%
        {IGPO}
\bibfield{author}{\bibinfo{person}{Guoqing Wang}, \bibinfo{person}{Sunhao Dai}, \bibinfo{person}{Guangze Ye}, \bibinfo{person}{Zeyu Gan}, \bibinfo{person}{Wei Yao}, \bibinfo{person}{Yong Deng}, \bibinfo{person}{Xiaofeng Wu}, {and} \bibinfo{person}{Zhenzhe Ying}.} \bibinfo{year}{2026}\natexlab{}.
\newblock \showarticletitle{Information Gain-based Policy Optimization: A Simple and Effective Approach for Multi-Turn Search Agents}. In \bibinfo{booktitle}{\emph{The Fourteenth International Conference on Learning Representations}}.
\newblock
\urldef\tempurl%
\url{https://openreview.net/forum?id=qkWP6phrvZ}
\showURL{%
\tempurl}


\bibitem[Wang et~al\mbox{.}(2025)]%
        {coRAG}
\bibfield{author}{\bibinfo{person}{Liang Wang}, \bibinfo{person}{Haonan Chen}, \bibinfo{person}{Nan Yang}, \bibinfo{person}{Xiaolong Huang}, \bibinfo{person}{Zhicheng Dou}, {and} \bibinfo{person}{Furu Wei}.} \bibinfo{year}{2025}\natexlab{}.
\newblock \bibinfo{title}{Chain-of-Retrieval Augmented Generation}.
\newblock
\showeprint[arxiv]{2501.14342}~[cs.IR]
\urldef\tempurl%
\url{https://arxiv.org/abs/2501.14342}
\showURL{%
\tempurl}


\bibitem[Wang et~al\mbox{.}(2022)]%
        {E5}
\bibfield{author}{\bibinfo{person}{Liang Wang}, \bibinfo{person}{Nan Yang}, \bibinfo{person}{Xiaolong Huang}, \bibinfo{person}{Binxing Jiao}, \bibinfo{person}{Linjun Yang}, \bibinfo{person}{Daxin Jiang}, \bibinfo{person}{Rangan Majumder}, {and} \bibinfo{person}{Furu Wei}.} \bibinfo{year}{2022}\natexlab{}.
\newblock \showarticletitle{Text embeddings by weakly-supervised contrastive pre-training}.
\newblock \bibinfo{journal}{\emph{arXiv}} (\bibinfo{year}{2022}).
\newblock
\urldef\tempurl%
\url{https://arxiv.org/abs/2212.03533}
\showURL{%
\tempurl}


\bibitem[Wang et~al\mbox{.}(2024)]%
        {RAT}
\bibfield{author}{\bibinfo{person}{Zihao Wang}, \bibinfo{person}{Anji Liu}, \bibinfo{person}{Haowei Lin}, \bibinfo{person}{Jiaqi Li}, \bibinfo{person}{Xiaojian Ma}, {and} \bibinfo{person}{Yitao Liang}.} \bibinfo{year}{2024}\natexlab{}.
\newblock \bibinfo{title}{RAT: Retrieval Augmented Thoughts Elicit Context-Aware Reasoning in Long-Horizon Generation}.
\newblock
\showeprint{2403.05313}


\bibitem[Webber et~al\mbox{.}(2010)]%
        {RBO}
\bibfield{author}{\bibinfo{person}{William Webber}, \bibinfo{person}{Alistair Moffat}, {and} \bibinfo{person}{Justin Zobel}.} \bibinfo{year}{2010}\natexlab{}.
\newblock \showarticletitle{A similarity measure for indefinite rankings}.
\newblock \bibinfo{journal}{\emph{ACM Trans. Inf. Syst.}} \bibinfo{volume}{28}, \bibinfo{number}{4}, Article \bibinfo{articleno}{20} (\bibinfo{date}{Nov.} \bibinfo{year}{2010}), \bibinfo{numpages}{38}~pages.
\newblock
\showISSN{1046-8188}
\href{https://doi.org/10.1145/1852102.1852106}{doi:\nolinkurl{10.1145/1852102.1852106}}


\bibitem[Xia et~al\mbox{.}(2026)]%
        {searchP1}
\bibfield{author}{\bibinfo{person}{Tianle Xia}, \bibinfo{person}{Ming Xu}, \bibinfo{person}{Lingxiang Hu}, \bibinfo{person}{Yiding Sun}, \bibinfo{person}{Wenwei Li}, \bibinfo{person}{Linfang Shang}, \bibinfo{person}{Liqun Liu}, \bibinfo{person}{Peng Shu}, \bibinfo{person}{Huan Yu}, {and} \bibinfo{person}{Jie Jiang}.} \bibinfo{year}{2026}\natexlab{}.
\newblock \bibinfo{title}{Search-P1: Path-Centric Reward Shaping for Stable and Efficient Agentic RAG Training}.
\newblock
\showeprint[arxiv]{2602.22576}~[cs.CL]
\urldef\tempurl%
\url{https://arxiv.org/abs/2602.22576}
\showURL{%
\tempurl}


\bibitem[Yang et~al\mbox{.}(2025a)]%
        {qwen25}
\bibfield{author}{\bibinfo{person}{An Yang}, \bibinfo{person}{Baosong Yang}, \bibinfo{person}{Beichen Zhang}, \bibinfo{person}{Binyuan Hui}, \bibinfo{person}{Bo Zheng}, \bibinfo{person}{Bowen Yu}, \bibinfo{person}{Chengyuan Li}, \bibinfo{person}{Dayiheng Liu}, \bibinfo{person}{Fei Huang}, \bibinfo{person}{Haoran Wei}, \bibinfo{person}{Huan Lin}, \bibinfo{person}{Jian Yang}, \bibinfo{person}{Jianhong Tu}, \bibinfo{person}{Jianwei Zhang}, \bibinfo{person}{Jianxin Yang}, \bibinfo{person}{Jiaxi Yang}, \bibinfo{person}{Jingren Zhou}, \bibinfo{person}{Junyang Lin}, \bibinfo{person}{Kai Dang}, \bibinfo{person}{Keming Lu}, \bibinfo{person}{Keqin Bao}, \bibinfo{person}{Kexin Yang}, \bibinfo{person}{Le Yu}, \bibinfo{person}{Mei Li}, \bibinfo{person}{Mingfeng Xue}, \bibinfo{person}{Pei Zhang}, \bibinfo{person}{Qin Zhu}, \bibinfo{person}{Rui Men}, \bibinfo{person}{Runji Lin}, \bibinfo{person}{Tianhao Li}, \bibinfo{person}{Tianyi Tang}, \bibinfo{person}{Tingyu Xia}, \bibinfo{person}{Xingzhang Ren},
  \bibinfo{person}{Xuancheng Ren}, \bibinfo{person}{Yang Fan}, \bibinfo{person}{Yang Su}, \bibinfo{person}{Yichang Zhang}, \bibinfo{person}{Yu Wan}, \bibinfo{person}{Yuqiong Liu}, \bibinfo{person}{Zeyu Cui}, \bibinfo{person}{Zhenru Zhang}, {and} \bibinfo{person}{Zihan Qiu}.} \bibinfo{year}{2025}\natexlab{a}.
\newblock \bibinfo{title}{Qwen2.5 Technical Report}.
\newblock
\showeprint[arxiv]{2412.15115}~[cs.CL]
\urldef\tempurl%
\url{https://arxiv.org/abs/2412.15115}
\showURL{%
\tempurl}


\bibitem[Yang et~al\mbox{.}(2025b)]%
        {simRAG}
\bibfield{author}{\bibinfo{person}{Diji Yang}, \bibinfo{person}{Linda Zeng}, \bibinfo{person}{Jinmeng Rao}, {and} \bibinfo{person}{Yi Zhang}.} \bibinfo{year}{2025}\natexlab{b}.
\newblock \showarticletitle{Knowing You Don't Know: Learning When to Continue Search in Multi-round RAG through Self-Practicing}. In \bibinfo{booktitle}{\emph{Proceedings of the 48th International ACM SIGIR Conference on Research and Development in Information Retrieval}} \emph{(\bibinfo{series}{SIGIR '25})}. \bibinfo{publisher}{Association for Computing Machinery}, \bibinfo{pages}{1305–1315}.
\newblock
\showISBNx{9798400715921}
\href{https://doi.org/10.1145/3726302.3730018}{doi:\nolinkurl{10.1145/3726302.3730018}}


\bibitem[Yang et~al\mbox{.}(2018)]%
        {hotpotQA}
\bibfield{author}{\bibinfo{person}{Zhilin Yang}, \bibinfo{person}{Peng Qi}, \bibinfo{person}{Saizheng Zhang}, \bibinfo{person}{Yoshua Bengio}, \bibinfo{person}{William~W. Cohen}, \bibinfo{person}{Ruslan Salakhutdinov}, {and} \bibinfo{person}{Christopher~D. Manning}.} \bibinfo{year}{2018}\natexlab{}.
\newblock \bibinfo{title}{HotpotQA: A Dataset for Diverse, Explainable Multi-hop Question Answering}.
\newblock
\showeprint[arxiv]{1809.09600}~[cs.CL]
\urldef\tempurl%
\url{https://arxiv.org/abs/1809.09600}
\showURL{%
\tempurl}


\bibitem[Zendel et~al\mbox{.}(2019)]%
        {informationNeed&Query&QPP}
\bibfield{author}{\bibinfo{person}{Oleg Zendel}, \bibinfo{person}{Anna Shtok}, \bibinfo{person}{Fiana Raiber}, \bibinfo{person}{Oren Kurland}, {and} \bibinfo{person}{J.~Shane Culpepper}.} \bibinfo{year}{2019}\natexlab{}.
\newblock \showarticletitle{Information Needs, Queries, and Query Performance Prediction}. In \bibinfo{booktitle}{\emph{Proceedings of the 42nd International ACM SIGIR Conference on Research and Development in Information Retrieval}} \emph{(\bibinfo{series}{SIGIR'19})}. \bibinfo{publisher}{Association for Computing Machinery}, \bibinfo{pages}{395–404}.
\newblock
\showISBNx{9781450361729}
\href{https://doi.org/10.1145/3331184.3331253}{doi:\nolinkurl{10.1145/3331184.3331253}}


\end{thebibliography}

\end{document}